\documentclass{article}
\usepackage{microtype}
\usepackage{graphicx}
\usepackage{subfigure}
\usepackage{booktabs}
\usepackage{hyperref}

\usepackage[accepted]{icml2024}

\usepackage{amsmath}
\usepackage{amssymb}
\usepackage{mathtools}
\usepackage{amsthm}
\usepackage[capitalize,noabbrev]{cleveref}
\usepackage{array}
\usepackage{booktabs}

\theoremstyle{plain}

\usepackage[textsize=tiny]{todonotes}

\usepackage{multirow}
\usepackage{float}
\usepackage{stfloats}
\usepackage{bbding}
\usepackage{pifont}
\renewcommand{\dblfloatpagefraction}{.95}
\renewcommand{\algorithmicrequire}{\textbf{Input:}}
\renewcommand{\algorithmicensure}{\textbf{Output:}}

\icmltitlerunning{Learning to Predict Mutation Effects of Protein-Protein Interactions by Microenvironment-aware Hierarchical Prompt Learning}

\begin{document}

\twocolumn[
\icmltitle{Learning to Predict Mutational Effects of Protein-Protein Interactions by Microenvironment-aware Hierarchical Prompt Learning}

\icmlsetsymbol{equal}{*}

\begin{icmlauthorlist}
\icmlauthor{Lirong Wu}{xxx}
\icmlauthor{Yijun Tian}{yyy}
\icmlauthor{Haitao Lin}{xxx}
\icmlauthor{Yufei Huang}{xxx}
\icmlauthor{Siyuan Li}{xxx}
\icmlauthor{Nitesh V Chawla}{yyy}
\icmlauthor{Stan Z. Li}{xxx}
\end{icmlauthorlist}

\icmlaffiliation{xxx}{Westlake University}
\icmlaffiliation{yyy}{University of Notre Dame}

\icmlcorrespondingauthor{Lirong Wu}{wulirong@westlake.edu.cn}

\icmlkeywords{Machine Learning, ICML}

\vskip 0.3in
]

\printAffiliationsAndNotice{}

\begin{abstract}
Protein-protein bindings play a key role in a variety of fundamental biological processes, and thus predicting the effects of amino acid mutations on protein-protein binding is crucial. To tackle the scarcity of annotated mutation data, pre-training with massive unlabeled data has emerged as a promising solution. However, this process faces a series of challenges: (1) complex higher-order dependencies among multiple (more than paired) structural scales have not yet been fully captured; (2) it is rarely explored how mutations alter the local conformation of the surrounding microenvironment; (3) pre-training is costly, both in data size and computational burden. In this paper, we first construct a hierarchical prompt codebook to record common microenvironmental patterns at different structural scales independently. Then, we develop a novel codebook pre-training task, namely masked microenvironment modeling, to model the joint distribution of each mutation with their residue types, angular statistics, and local conformational changes in the microenvironment. With the constructed prompt codebook, we encode the microenvironment around each mutation into multiple hierarchical prompts and combine them to flexibly provide information to wild-type and mutated protein complexes about their microenvironmental differences. Such a hierarchical prompt learning framework has demonstrated superior performance and training efficiency over state-of-the-art pre-training-based methods in mutation effect prediction and a case study of optimizing human antibodies against SARS-CoV-2.
\end{abstract}

\vspace{-2.5em}
\section{Introduction}
\vspace{-0.5em}
Proteins usually interact with other proteins to perform specific biological functions that are essential for all organisms~\cite{hu2021survey,kastritis2013binding,lu2020recent,gao2024proteininvbench,lin2024functional,huang2024protein,wu2024psc}. A prime example is antibodies, a family of Y-shape proteins produced by the immune system to recognize, bind, and interact with proteins on the surface of pathogens~\cite{murphy2016janeway,tan2024cross}. Therefore, how to develop methods to modulate protein-protein interactions has become a key issue, and one of the most prevalent strategies is to mutate the amino acids at the interaction interface~\cite{luo2023rotamer,liu2023predicting}. Considering the enormous combinatorial space of over $20^{30}$ amino acid mutations and the high variability of mutated structures, it is not feasible to test all potential mutations by experimental assays in the web laboratory, which calls for computational methods to screen for desirable mutations by predicting binding affinity changes of protein complexes upon mutations. This problem, also known as \emph{the change in binding free energy} ($\Delta\Delta G$) prediction, is a core challenge in the protein complex design~\cite{marchand2022computational}.

\vspace{-0.15em}
The computational methods for $\Delta\Delta G$ prediction have undergone a paradigm shift from biophysics-based and statistics-based techniques~\cite{schymkowitz2005foldx,park2016simultaneous,alford2017rosetta} to Deep Learning (DL) techniques~\cite{shan2022deep,luo2023rotamer,liu2023predicting}. Despite the great progress made by DL-based methods, the scarcity of annotated experimental data and the unavailability of mutated complex structures remain two major challenges for effective supervised learning. Therefore, pre-training with massive unlabeled data is becoming a promising solution. On the one hand, the general knowledge learned from the pre-training data can be transferred for $\Delta\Delta G$ prediction, which solves the problem of data sparsity effectively. On the other hand, some of the pre-training tasks can capture sequence-structure dependencies, enabling the model to be implicitly aware of the mutated complex structures rather than explicitly predicting them. Owing to these two benefits, pre-training is becoming one of the most prevalent strategies for $\Delta\Delta G$ prediction~\cite{luo2023rotamer}.

\vspace{-0.15em}
Despite the fruitful progress, existing pre-training-based methods still encounter several key issues. The first is the ignorance of modeling multiple types of structural scales and their dependencies. A protein can focus on different structural scales to implement specific functions, and each structural scale has its own merits and cannot replace each other Besides, the dependencies between different structural scales are diverse, and simply focusing on single or paired structural scales using existing pre-training tasks cannot fully capture their complex higher-order dependencies. The second obstacle is the lack of mutated complex structures. Although Alphafold2 (AF2)~\cite{jumper2021highly} and ESMFold~\cite{lin2023evolutionary} have made great advances in protein structure prediction, they still struggle to predict the exact conformational changes upon subtle mutations in amino acids. Moreover, it has been found that the performance of a model trained with experimental structures drops significantly when tested on the predicted AF2 structures~\cite{huang2023data}. As an alternative, Rotamer Density Estimator (RDE)~\cite{luo2023rotamer} and DiffAffinity~\cite{liu2023predicting} implicitly model sequence-structure dependence by predicting \emph{global} sidechain conformational changes upon \emph{all mutations}, ignoring how \emph{each mutation} alter the \emph{local} backbone conformation of it surrounding microenvironment. Moreover, the computational cost in existing pre-training tasks caused by the huge amount of data is too expensive and even far beyond the task of $\Delta\Delta G$ prediction itself. For example, there are only 7k labeled mutation data in the SKEMPI v2.0 dataset, but the PDB-REDO dataset used for pre-training by RDE contains more than 143k data.

\begin{figure}[!tbp]
    \vspace{-0.5em}
	\begin{center}
		\includegraphics[width=0.49\linewidth]{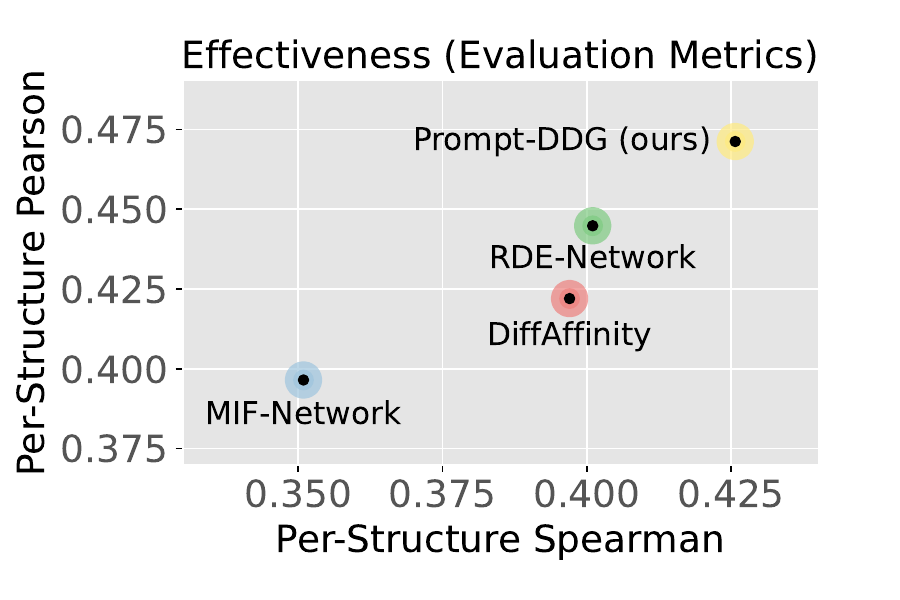}
		\includegraphics[width=0.49\linewidth]{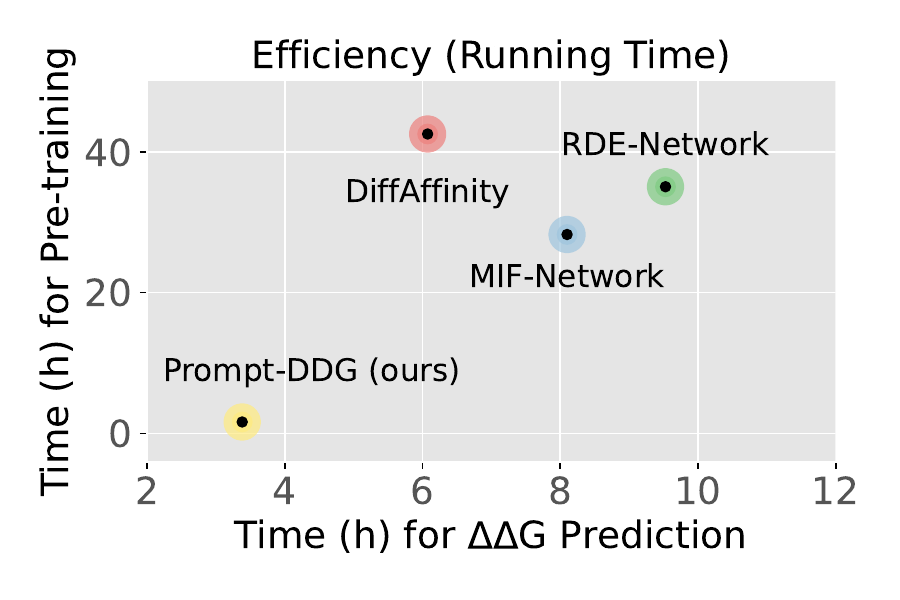}
	\end{center}
    \vspace{-1.5em}
	\caption{Comparison of our Prompt-DDG with three state-of-the-art methods in effectiveness (per-structure Pearson and Spearman) and training efficiency (for pre-training and $\Delta\Delta G$ prediction). where Prompt-DDG outperforms the other methods a lot in both effectiveness and efficiency, especially the time spent on pre-training.}
	\label{fig:1}
    \vspace{-1em}
\end{figure}

\textbf{Present Work.} In this paper, we propose a simple yet effective \textit{Microenvironment-aware Hierarchical Prompt Learning} framework for efficient $\Delta\Delta G$ prediction (Prompt-DDG). The core idea of Prompt-DDG is to avoid the computationally heavy pre-training and instead directly generate concise prompts for each mutation in an \emph{lightweight and efficient} manner. These prompts are expected to characterize the microenvironmental differences surrounding the mutation between wild-type and mutated complexes. To enable the generated prompts to fully cover the diversity of structural scales of the microenvironment, we construct a hierarchical prompt codebook to separately record common microenvironmental patterns of different structural scales. A novel codebook pre-training task, namely masked microenvironment modeling, is then proposed to model the joint distribution of each residue mutation and their heterogeneous properties, including residue types, angular statistics, and local conformational changes in the microenvironment. Using the hierarchical prompt codebook, we encode the microenvironment around each mutation into several prompts, which are passed through a lightweight module to flexibly provide wild-type and mutated complexes with multi-scale structural information about their microenvironments. Finally, Prompt-DDG outperforms other leading methods in terms of both effectiveness and efficiency, as shown in Figure.~\ref{fig:1},

\vspace{-1.0em}
\section{Related Work}
\vspace{-0.4em}
\subsection{Mutation Effect Prediction For Single Proteins}
\vspace{-0.4em}
The prediction of mutation effects for single proteins is mainly aimed at predicting changes in the stability, fluorescence, fitness, or other properties of proteins upon the mutations~\cite{alford2017rosetta,lei2023mutational,meier2021language}. The current mainstream is mainly sequence-based methods, which exploit co-evolutionary information mined by Multiple Sequence Alignments (MSAs)~\cite{frazer2021disease,luo2021ecnet} or Protein Language Models (PLMs)~\cite{meier2021language,notin2022tranception}. However, these methods are difficult to directly extend for the prediction of mutation effects on protein-protein interactions. For one thing, protein complexes involve multiple proteins or chains that may belong to different species and thus lack co-evolutionary information~\cite{luo2023rotamer}. Secondly, it is more difficult to predict changes in the binding free energy between proteins upon mutations than changes in the functions of single proteins. Finally, protein-protein interactions are mainly determined by protein structure than sequence. Therefore, mutation effect prediction on PPIs requires more efficient use of protein structures, rather than only protein sequences.

\begin{figure*}[!tbp]
    \vspace{-0.8em}
	\begin{center}
		\includegraphics[width=1.0\linewidth]{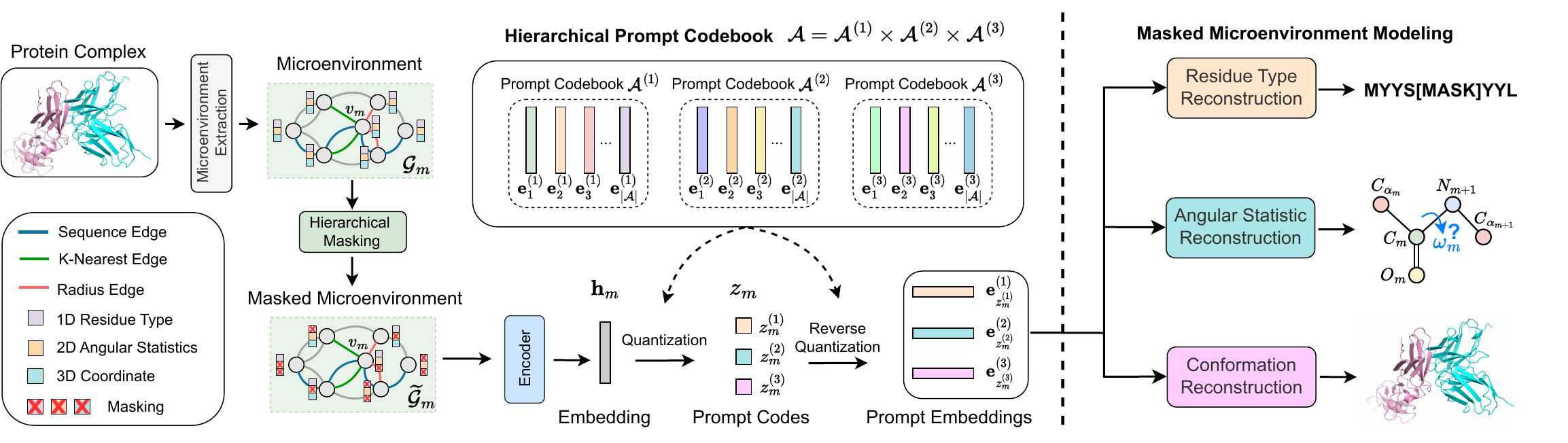}
	\end{center}
    \vspace{-1.2em}
	\caption{Left: A high-level overview of microenvironment-aware hierarchical prompt learning and adaptation framework for efficient $\Delta\Delta G$ prediction (Prompt-DDG). Right: Illustration of a hierarchical pre-training task by Masked Microenvironment Modeling (MMM).}
    \vspace{-1em}
	\label{fig:2}
\end{figure*}

\vspace{-0.5em}
\subsection{Mutation Effect Prediction For Protein Complexes}
\vspace{-0.5em}
Traditional approaches for predicting the effects of mutations on protein-protein binding ($\Delta\Delta G$) can be mainly divided into two branches: biophysics-based and statistics-based methods. The biophysics-based~\cite{alford2017rosetta, park2016simultaneous,delgado2019foldx} approaches sample mutated conformations from the energy function and then simulate inter-atomic interactions for predicting $\Delta\Delta G$, and thus face a trade-off between efficiency and effectiveness. On the other hand, statistics-based approaches~\cite{geng2019isee,li2016mutabind} use geometrical, physical, and evolutionary descriptors of proteins to predict mutational effects, and are thus limited by the choice of descriptors and cannot leverage the growing availability of protein structures.

\vspace{-0.1em}
Recent deep learning-based approaches can be categorized into two classes: end-to-end and pre-training-based methods. The end-to-end approaches~\cite{shan2022deep,luo2023rotamer} train a feature encoder to extract representations of both wild-type and mutated protein complexes, and then combine the two to directly predict $\Delta\Delta G$. The pre-training-based approaches learn sequence-structure dependencies by pre-training on large amounts of unlabeled data and then transfer the learned knowledge for predicting $\Delta\Delta G$. For example, Masked Inverse Folding (MIF)~\cite{yang2022masked} treats protein inverse folding as a pretext task to learn deep transformations from structure to sequence. Instead, RDE~\cite{luo2023rotamer} employs normalizing flows to estimate the distribution of protein side-chain conformations and then uses entropy to measure flexibility. Similarly, DiffAffinity~\cite{liu2023predicting} pre-trains a side-chain diffusion probabilistic model on unlabeled protein structures and leverages the pre-trained representations to predict $\Delta\Delta G$.

\vspace{-1em}
\section{Preliminary}
\vspace{-0.5em}
\textbf{Graph Construction.} We represent each protein-protein complex as a \textit{Heterogeneous Attribute Graph} $\mathcal{G}=(\mathcal{V}, \mathcal{E})$, where $\mathcal{V}=(\mathcal{V}_L, \mathcal{V}_R)$ is the node (residue) set of the ligand $\mathcal{V}_L$ and the receptor $\mathcal{V}_R$, and $\mathcal{E}=(\mathcal{E}_{\text{in}}, \mathcal{E}_{\text{ex}})$ separately contain internal edges within each component and external edges between components. Each node $v_i=(\mathbf{x}_i, \mathbf{Z}_i)\in\mathcal{V}$ is attributed as a node feature vector $\mathbf{x}_i\in\mathbb{R}^{d_n}$ and a node coordinate matrix $\mathbf{Z}_i\in\mathbb{R}^{3\times 4}$ consisting of 4 backbone atoms $\{N, C_{\alpha}, C, O\}$. In addition, each edge $e_{i,j}\in\mathcal{E}$ is described by an edge feature vector $\mathbf{E}_{i,j}\in\mathbb{R}^{d_e}$. Therefore, the graph of each protein-protein complex can also be denoted as $\mathcal{G}=(\mathbf{X}, \mathbf{Z}, \mathbf{E})$. For each node $v_i\in\mathcal{V}$, we define its node feature $\mathbf{x}_i$ as E(3)-invariant feature, as follows 
\vspace{-0.3em}
\begin{equation}
\begin{small}
\begin{aligned}
    \mathbf{x}_i=\Big\{E_{\text{type}}(v_i), E_{\text{ang}}(\Omega_i), Q_i^{\top}\frac{Z_{i,\xi}\!-\!Z_{i,C_\alpha}}{\left\|Z_{i,\xi}\!-\!Z_{i,C_\alpha}\right\|} \ \big|\  \Omega_i, \xi\Big\},
\end{aligned}
\end{small}
\end{equation}
where $E_{\text{type}}(v_i)$ is trainable type embedding of residue $v_i$. $\Omega_i$ contains three dihedral angles $\alpha_i,\beta_i,\gamma_i$ of the backbone and four torsion angles $\{\chi_{i}^{(k)}\}_{k=1}^{4}$ of the side chain, and $E_{\text{ang}}(\cdot)$ denotes the angular encodings~\cite{luo2023rotamer} in $\Omega_i$. The last term in $\mathbf{x}_i$ is direction encodings that correspond to the relative directions of three backbone atoms $\xi\in\{C,N,O\}$ in the local coordinate frame $Q_i$ of residue $v_i$. The edge feature $\mathbf{E}_{i,j}$ that describes the relationship between two residues $v_i$ and $v_j$ is defined as follows
\vspace{-0.3em}
\begin{equation}
\begin{small}
\begin{aligned}
\hspace{-1.2em}
\mathbf{E}_{i,j}\!=\!\Big\{E_{\text{pos}}(i,j), E_{\text{dis}}(\mathbf{Z}_i,\mathbf{Z}_j),  Q_i^{\top}\frac{Z_{j,\zeta}\!-\!Z_{i,C_\alpha}}{\left\|Z_{j,\zeta}\!-\!Z_{i,C_\alpha}\right\|} \big| \ \zeta\Big\},
\end{aligned}
\end{small}
\end{equation}
where $E_{\text{pos}}(i,j)$ and $E_{\text{dis}}(\mathbf{Z}_i,\mathbf{Z}_j)$ encode the relative sequential and spatial distances between residue $v_i$ and residue $v_j$, respectively. $E_{\text{pos}}(i,j)$ is set as 0 for any external edge $e_{i,j}\in\mathcal{E}_{\text{ex}}$. In addition, the last term is the direction encodings of four backbone atoms $\zeta\in\{C_\alpha, C,N,O\}$ of residue $v_j$ in the local coordinate frame $Q_i$ of residue $v_i$.

\textbf{Problem Statement.} Given a wild-type protein complex $\mathcal{G}^W\!=\!(\mathbf{X}, \mathbf{Z}, \mathbf{E})$ and a set of mutations $\mathcal{M}$, the task of mutational effect prediction for protein complexes can be formulated as predicting the change in $\Delta\Delta G$ between the wild-type complex $\mathcal{G}^W$ and mutated complex $\mathcal{G}^M\!=\!g(\mathcal{G}^W, \mathcal{M})$, i.e., approximating the mapping $p(\Delta\Delta G \mid \mathcal{G}^W, \mathcal{G}^M)$.

\vspace{-0.5em}
\section{Methodology}
\vspace{-0.5em}
In this paper, we propose a Prompt-DDG framework with three novel components for $\Delta\Delta G$ prediction. The pipeline is shown in Figure.~\ref{fig:2}. In particular, the first component constructs a hierarchical prompt codebook that encodes the microenvironment around each mutation as prompts of different structural scales. The second component pre-trains the prompt codebook hierarchically by masked microenvironment modeling. The third component adopts a lightweight prompt adaptation module that combines prompts of different scales to provide the microenvironmental differences.

\vspace{-0.6em}
\subsection{Hierarchical Prompt Codebook Construction}
\vspace{-0.4em}
\subsubsection{Definition of Microenvironment}
\vspace{-0.2em}
The microenvironment of a residue describes its surrounding sequence and structure contexts. We follow~\cite{wu2024mapeppi} to define the microenvironment of each mutation $v_m\in\mathcal{M}$ as a $v_m$-ego subgraph $\mathcal{G}_m\subseteq\mathcal{G}$ of the protein complex graph $\mathcal{G}$, with its node set $V_{\mathcal{G}_m}$ defined as follows
\begin{equation*}
\begin{small}
\begin{aligned}
\hspace{-1.2em}
    V_{\mathcal{G}_m} \!=\! \left\{v_n \ \big|\  |m\!-\!n|\!\leq\! d_s, \big\|Z_{m,C_\alpha}\!-\!Z_{n,C_\alpha}\big\|\!\leq\! d_r, v_n\!\in\!\mathcal{N}^{(K)}_m\right\}, 
\end{aligned}
\end{small}
\end{equation*}
where $d_s$ and $d_r$ are cut-off distances, $Z_{m,C_\alpha}$ and $Z_{n,C_\alpha}$ are the 3D coordinates of carbon-alpha atoms, and $\mathcal{N}^{(K)}_m$ is the $K$-hop neighborhood of residue $v_m$ in the spatial space. 

\vspace{-0.5em}
\subsubsection{Hierarchical Codebook Construction}
\vspace{-0.2em}
Protein-protein interactions focus on different structural scales to implement their functions, such as 1D for amino acid sequences, 2D geometric statistics, 3D structural coordinates, etc. Each structural scale has its own merits and cannot replace each other. To fully capture the different structural scales of the microenvironment, we constructed a hierarchical prompt codebook $\mathcal{A}=\mathcal{A}^{(1)}\!\times\!\mathcal{A}^{(2)}\!\times\!\mathcal{A}^{(3)}$, where $\mathcal{A}^{(1)}$, $\mathcal{A}^{(2)}$, and $\mathcal{A}^{(3)}$ characterize three aspects of the microenvironment, i.e., residue type, angular statistics, and local conformation, respectively. Each sub-codebook $\mathcal{A}^{(k)}$ is parameterized as $\mathcal{A}^{(k)}=\{\mathbf{e}_1^{(k)},\mathbf{e}_2^{(k)},\cdots,\mathbf{e}_{|\mathcal{A}|}^{(k)}\}\in\mathbb{R}^{|\mathcal{A}|\times F}$, where $\{\mathbf{e}_{i}^{(k)}\}_{i=1}^{|\mathcal{A}|}$ are $|\mathcal{A}|$ learnable prompt embeddings. 

To generate microenvironment-aware prompts for different structural scales, we first encode the microenvironment $\mathcal{G}_m$ into a hidden representation $\mathbf{h}_m$ using a self-attention-based graph neural network $f_\theta(\cdot)$ that is invariant to rotation and translation~\cite{jumper2021highly}. Next, the microenvironments $\{\mathcal{G}_1,\mathcal{G}_2,\cdots,\mathcal{G}_{|\mathcal{M}|}\}$ of $|\mathcal{M}|$ mutations can be tokenized to discrete prompt codes $\{z_1,z_2,\cdots,z_M\}$ by vector quantization \citep{van2017neural} that looks up the nearest neighbors in the hierarchical codebook $\mathcal{A}$. For each microenvironment $\mathcal{G}_m$, its prompt codes $z_i=\{z_m^{(1)},z_m^{(2)},z_m^{(3)}\}$ can be defined as follows
\begin{equation}
z_m^{(k)}=\operatorname{argmin}_n\big\|\mathbf{h}_m-\mathbf{e}^{(k)}_n\big\|_2,\ \ \text{where} \ 1\leq k \leq 3.
\label{equ:3}
\end{equation}

\vspace{-1.2em}
\subsection{Masked Microenvironment Modeling (MMM)}
\vspace{-0.2em}
To resolve the non-differentiability of the vector quantization, we impose a constraint $\mathcal{L}_{\text{VQ}}$ to bride the codebook $|\mathcal{A}|$ and microenvironment representations by straight-through estimator \citep{bengio2013estimating}, which is defined as follows
\vspace{-0.5em}
\begin{equation}
\hspace{-0.8em}
\begin{aligned}
\mathcal{L}_{\text{VQ}}\!=\!\frac{1}{|\mathcal{V}|}\sum_{v_i\in\mathcal{V}}\sum_{k=1}^3\Big(\big\|\mathrm{sg}\big[\mathbf{h}_i\big]\!-\!\mathbf{e}_{z_i^{(k)}}\big\|_2^2 \!+\!\eta\big\|\mathbf{h}_i\!-\!\mathrm{sg}\big[\mathbf{e}_{z_i^{(k)}}\big]\big\|_2^2\Big),
\end{aligned}
\label{equ:4}
\end{equation}
where $\eta$ is a trade-off hyperparameter, and $\mathrm{sg}[\cdot]$ is the stop-gradient operation. The first term in Eq.~(\ref{equ:4}) is a codebook loss, used to update the codebook to make the microenvironment representations $\mathbf{h}_i$ close to the most similar prompt embeddings. The second term in Eq.~(\ref{equ:4}) is a commitment loss, encouraging the encoder outputs to stay close to the chosen prompt embeddings by only training the encoder.

Next, we focus on how to pre-train learnable prompt embeddings in the constructed hierarchical codebook $\mathcal{A}$. To this end, we take three different data reconstruction tasks to simultaneously learn the hierarchical codebook $\mathcal{A}$ and train the microenvironment encoder $f_\theta(\cdot)$. We train each sub-codebook hierarchically with individual reconstruction tasks to make it focus on a specialized structural scale. Furthermore, in order to fully capture the higher-order (more than single and paired) dependencies among various structural scales, we unify the hierarchical training in one pre-training task, i.e., masked microenvironment modeling. Specifically, we independently mask the residue types, geometric angles, and conformation coordinates in the microenvironment $\mathcal{G}_m$ by randomly flipping, zeroing, and Gaussian noising, and then reconstruct the inputs from the masked microenvironment $\widetilde{\mathcal{G}}_m$ by three different reconstruction tasks. The masked residues sets of three structural scales are independent and are denoted as $\mathcal{C}_1$, $\mathcal{C}_2$, and $\mathcal{C}_3$, respectively.

\textbf{Residue Type Reconstruction.}
To pre-train the sub-codebook $\mathcal{A}^{(1)}$, we use a symmetric variant of microenvironment encoder $f_\theta(\cdot)$ as the type decoder $\widehat{f}_\theta(\cdot)$, which predicts the residue types $\{\widehat{s}_i\}_{v_i\in\mathcal{C}_1}$ of a masked residue set $\mathcal{C}_1\subseteq\mathcal{V}$ from the prompt embeddings $\{\mathbf{e}_{z_i^{(1)}}\}_{v_i\in\mathcal{V}}$. The reconstruction loss $\mathcal{L}_{\text {seq}}$ is defined by cross-entropy $\ell_{c e}$:
\vspace{-0.5em}
\begin{equation}
\mathcal{L}_{\text {seq }}\big(\mathcal{A}^{(1)}\big)=\frac{1}{|\mathcal{C}_1|} \sum_{v_i\in\mathcal{C}_1} \ell_{c e}\big(s_i, \widehat{s}_i\big).
\end{equation}

\vspace{-0.8em}
\textbf{Angular Statistic Reconstruction.} To pre-train the sub-codebook $\mathcal{A}^{(2)}$, we use another decoder to reconstruct the angular information from the prompt embeddings $\{\mathbf{e}_{z_i^{(2)}}\}_{v_i\in\mathcal{V}}$. Next, we compute the MSE loss between the predicted and the ground-truth ones as the loss $\mathcal{L}_{\text {ang}}$:
\begin{equation}
\mathcal{L}_{\text {ang}}\big(\mathcal{A}^{(2)}\big)=\frac{1}{|\mathcal{C}_2|} \sum_{v_i\in\mathcal{C}_2} \sum_{a\in\Omega_i} \Big\| E_{\text{ang}}(a) - E_{\text{ang}}(\widehat{a}) \Big\|_2^2,
\end{equation}
where $\mathcal{C}_2\subseteq\mathcal{V}$ is the residue set with masked angles, $\Omega_i$ contains three dihedral angles $\alpha_i,\beta_i,\gamma_i$ of the backbone and four torsion angles $\{\chi_{i}^{(k)}\}_{k=1}^{4}$ of the side chain, and $E_{\text{ang}}(\cdot)$ denotes the angular encodings used by~\cite{luo2023rotamer}.

\textbf{Local Conformation Reconstruction.}
As mentioned earlier, a key issue for $\Delta\Delta G$ prediction is how to be aware of local conformational changes induced by mutations. Given the absence of mutated structures renders supervised learning infeasible, we take another unsupervised noise estimation task as a pretext task for pre-training the sub-codebook. Specifically, we first add Gaussian noise $\mathbf{O}_i$ to the wild-type structure $\mathbf{Z}_i$ to get noisy structure $\widetilde{\mathbf{Z}}_i=\mathbf{Z}_i+\mathbf{O}_i$, which is encoded by an E(3)-equivariant graph neural network (see \textbf{Appendix A} for a detailed description of the architecture) to predict structural noise $\widehat{\mathbf{O}}_i$, e.g., local conformation change, from the noisy structure $\widetilde{\mathbf{Z}}_i$ to the wild-type structure $\mathbf{Z}_i$. Furthermore, we adopt the Huber loss~\cite{huber1992robust} (see \textbf{Appendix B} for detailed formulas) other than the common MSE loss as the objective function, defined as follows:
\vspace{-0.5em}
\begin{equation}
\mathcal{L}_{\text {struct }}\big(\mathcal{A}^{(3)}\big)=\frac{1}{\left|\mathcal{C}_3\right|} \sum_{v_i \in \mathcal{C}_3} \ell_{\text {huber }}\left(\mathbf{O}_i, \widehat{\mathbf{O}}_i\right),
\vspace{-0.5em}
\end{equation}
where $\mathcal{C}_3\subseteq\mathcal{V}$ is the residue set with Gaussian noise added.

\textbf{Summary.} While each sub-codebook $\mathcal{A}^{(k)}$ ($1 \leq k \leq 3$)  is hierarchically trained with one individual reconstruction task, they share the same masked inputs and microenvironment encoder, which makes \emph{the reconstruction of individual structural scale dependent not only on itself but also on other scales}. Thus, MMM well models the joint distribution of each mutation with their residue types, angular statistics, and local conformation changes in the microenvironment.

\subsection{Prompt Training, Adaptation, and Usage}
\vspace{-0.3em}
\subsubsection{Lightweight Prompt Adaptation}
\vspace{-0.3em}
The final loss function $\mathcal{L}$ for the hierarchical training of the prompt codebook $\mathcal{A}=\mathcal{A}^{(1)}\!\times\!\mathcal{A}^{(2)}\!\times\!\mathcal{A}^{(3)}$ is defined as:
\begin{equation}
\mathcal{L} \!=\! \lambda\mathcal{L}_{\text {VQ}} \!+\! \mathcal{L}_{\text {seq}}\big(\mathcal{A}^{(1)}\big) \!+\! \mathcal{L}_{\text {ang}}\big(\mathcal{A}^{(2)}\big) \!+\! \mathcal{L}_{\text {struct}}\big(\mathcal{A}^{(3)}\big).
\label{equ:8}
\end{equation}
Using the pre-trained codebook $\mathcal{A}$, we can encode the microenvironment $\mathcal{G}_m$ around each mutation $m\in\mathcal{M}$ into three discrete prompt codes $z_i=\{z_m^{(1)},z_m^{(2)},z_m^{(3)}\}$ by Eq.~(\ref{equ:3}). Since different sub-codebooks record different structural scales of the microenvironment, we flexibly combine the acquired prompt embeddings $\{\mathbf{e}_{z_m^{(1)}},\mathbf{e}_{z_m^{(2)}},\mathbf{e}_{z_m^{(3)}}\}$ to bridge the gap between pre-trained prompts and downstream tasks. The prompt combination is implemented by a lightweight prompt adaptation module, defined as
\begin{equation}
\begin{aligned}
 \mathbf{p}_m =  \alpha_1\cdot\mathbf{e}_{z_m^{(1)}} & + \alpha_2\cdot\mathbf{e}_{z_m^{(2)}} + \alpha_3\cdot\mathbf{e}_{z_m^{(3)}},\ \text{where} \\
\alpha_k = & \phi_{\omega}^{(k)}\big(\mathbf{e}_{z_m^{(1)}}, \mathbf{e}_{z_m^{(2)}}, \mathbf{e}_{z_m^{(3)}}\big),
\end{aligned}
\label{equ:9}
\end{equation}
where $\{\phi_{\omega}^{(k)}(\cdot)\}_{k=1}^3$ are one-layer linear transformation.

\vspace{-0.5em}
\subsubsection{Prompt-Guided $\Delta\Delta G$ Prediction}
\begin{figure}[!tbp]
	\begin{center}
		\includegraphics[width=1.0\linewidth]{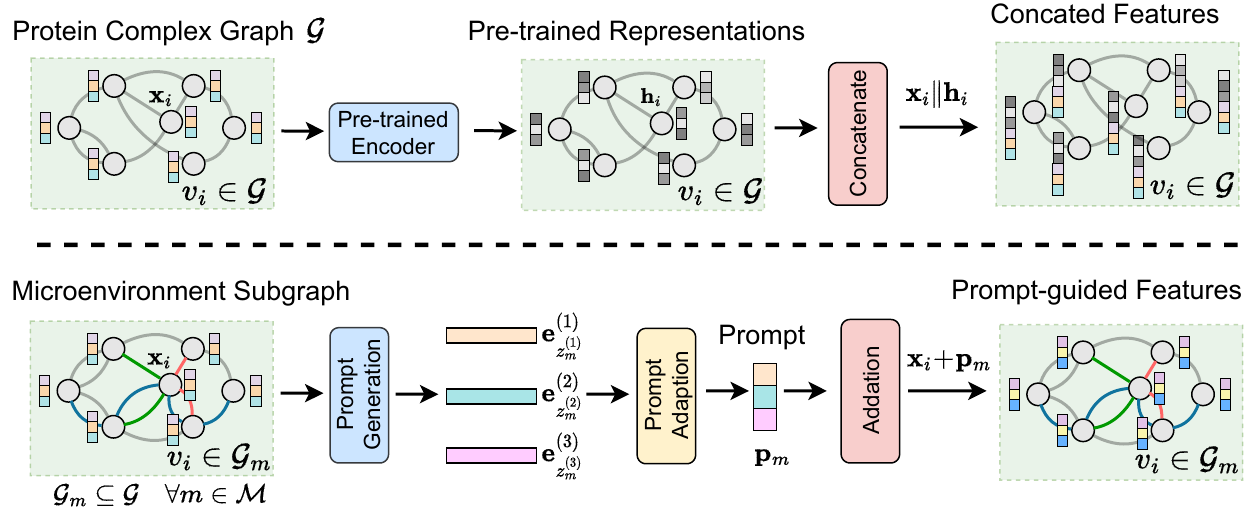}
	\end{center}
    \vspace{-1em}
	\caption{Top (pre-training-based): concat pre-trained representation $\mathbf{h}_i$ of residue $v_i\in\mathcal{V}$ with the original feature $\mathbf{x}_i$. Below (prompt-guided): encode the microenvironment $\mathcal{G}_m$ around mutation $m\in\mathcal{M}$ into a prompt $\mathbf{p}_m$ and add it to each residue $v_i\in\mathcal{G}_m$.}
    \vspace{-0.5em}
	\label{fig:3}
\end{figure}

\vspace{-0.3em}
Previous pre-training-based methods learn a pre-trained representation $\mathbf{h}_i$ for each residue $v_i\in\mathcal{V}$ and concat (or add) it to the corresponding residue, i.e., $\widetilde{\mathbf{x}}_i = \mathbf{x}_i \| \mathbf{h}_i$. In contrast, we encode the microenvironment $\mathcal{G}_m$ around each mutation $m\in\mathcal{M}$ as a prompt embedding $\mathbf{p}_m$ and then add it to each residue $v_i\in\mathcal{V}_{\mathcal{G}_m}$ within the microenvironment, i.e., $\widetilde{\mathbf{x}}_i = \mathbf{x}_i \!+\! \mathbf{p}_m$, as shown in Figure.~\ref{fig:3}. Next, we use a network that shares the same architecture as the microenvironment encoder $f_\theta(\cdot)$ to transform prompt-guided inputs $\{\widetilde{\mathbf{x}}_i\}_{v_i\in\mathcal{V}}$ and apply max-pooling to obtain a global structure representation. We then subtract the wild-type representation from the mutant representation, feed it into an MLP to predict $\Delta\Delta \widehat{G}$, and calculate the MSE loss between it and the ground-truth $\Delta\Delta G$ as the final objective function.

\vspace{-0.5em}
\subsection{Further Comparison and Discussion}
\vspace{-0.5em}
Compared to pre-training, prompt learning on proteins is still a new topic, and the only similar work so far is PromptProtein~\cite{wang2023multilevel}, but Prompt-DDG differs from it in four aspects: (1) PromptProtein focuses only on the task of property prediction for \emph{single proteins}, and whether and how it can be extended to \emph{protein complexes} like Prompt-DDG remains unexplored. (2) PromptProtein uses a \emph{learnable} attention mask matrix to model the relationship between prompts and existing residues, but prompts in Prompt-DDG are \emph{explicitly associated} with a specific microenvironment. (3) PromptProtein learns \emph{global task-specific} prompts for all residues, while Prompt-DDG learns a \emph{local mutation-specific} prompt for each residue within the microenvironment. (4) PromptProtein learns \emph{continuous} prompts, but Prompt-DDG constructs a hierarchical prompt codebook to record only those most common microenvironmental patterns in a \emph{discrete} fashion. In addition, another topic related to Prompt-DDG is the protein microenvironment encoding, such as a recent work MAPE-PPI~\cite{wu2024mapeppi}, which has been discussed in detail in \textbf{Appendix C}. Due to space limitations, the time complexity analysis and pseudo-code of our Prompt-DDG are available in \textbf{Appendix D \& E}.

\begin{table*}[!htbp]
\begin{center}
\vspace{-1em}
\caption{Mean results of 3-fold cross-validation on the SKEMPI v2 dataset, where \textbf{bold} and \underline{underline} denote the best and second metrics.}
\label{tab:1}
\resizebox{\textwidth}{!}{
\begin{tabular}{cl|cc|ccccc}

\toprule
\multirow{2}{*}{\textbf{Category}} & \multirow{2}{*}{\textbf{Method}} & \multicolumn{2}{c}{\textbf{Per-Structure}} & \multicolumn{5}{c}{\textbf{Overall}} \\ \cmidrule(r){3-4} \cmidrule(r){5-9}
 &  & \textbf{Pearson} $\uparrow$ & \textbf{Spear.} $\uparrow$ & \textbf{Pearson} $\uparrow$ & \textbf{Spear.} $\uparrow$ & \textbf{RMSE} $\downarrow$ & \textbf{MAE} $\downarrow$ & \textbf{AUROC} $\uparrow$ \\ \midrule
\multirow{4}{*}{Sequence-based} & ESM-1v & 0.0073 & -0.0118 & 0.1921 & 0.1572 & 1.9609 & 1.3683 & 0.5414 \\
 & PSSM & 0.0826 & 0.0822 & 0.0159 & 0.0666 & 1.9978 & 1.3895 & 0.5260 \\
 & MSA Transformer & 0.1031 & 0.0868 & 0.1173 & 0.1313 & 1.9835 & 1.3816 & 0.5768 \\
 & Tranception & 0.1348 & 0.1236 & 0.1141 & 0.1402 & 2.0382 & 1.3883 & 0.5885 \\ \midrule
\multirow{2}{*}{Energy Function} & Rosetta & 0.3284 & 0.2988 & 0.3113 & 0.3468 & 1.6173 & 1.1311 & 0.6562 \\
 & FoldX & 0.3789 & 0.3693 & 0.3120 & 0.4071 & 1.9080 & 1.3089 & 0.6582 \\ \midrule
\multirow{2}{*}{Supervised} & DDGPred & 0.3750 & 0.3407 & 0.6580 & 0.4687 & \textbf{1.4998} & \underline{1.0821} & 0.6992 \\
 & End-to-End & 0.3873 & 0.3587 & 0.6373 & 0.4882 & 1.6198 & 1.1761 & 0.7172 \\ \midrule
\multirow{7}{*}{Pre-training-based} & B-factor & 0.2042 & 0.1686 & 0.2390 & 0.2625 & 2.0411 & 1.4402 & 0.6044 \\
 & ESM-1F & 0.2241 & 0.2019 & 0.3194 & 0.2806 & 1.8860 & 1.2857 & 0.5899 \\
 & MIF-$\Delta$logit & 0.1585 & 0.1166 & 0.2918 & 0.2192 & 1.9092 & 1.3301 & 0.5749 \\
 & MIF-Network & 0.3965 & 0.3509 & 0.6523 & 0.5134 & 1.5932 & 1.1469 & 0.7329 \\
 & RDE-Linear & 0.2903 & 0.2632 & 0.4185 & 0.3514 & 1.7832 & 1.2159 & 0.6059 \\
 & RDE-Network & \underline{0.4448} & \underline{0.4010} & 0.6447 & \underline{0.5584} & 1.5799 & 1.1123 & \underline{0.7454} \\
 & DiffAffinity & 0.4220 & 0.3970 & \underline{0.6690} & 0.5560 & 1.5350 & 1.0930 & 0.7440 \\ \midrule
 \multirow{3}{*}{Ours} & Prompt-DDG & \textbf{0.4712} & \textbf{0.4257} & \textbf{0.6772} & \textbf{0.5910} & \underline{1.5207} & \textbf{1.0770} & \textbf{0.7568} \\
 & $\Delta_{\text{RDE-Network}}$ & +5.94\% & +6.16\% & +5.04\% & +5.84\% & +3.74\% & +3.17\% & +1.53\% \\
 & $\Delta_{\text{DiffAffinity}}$ & +11.78\% & +7.23\% & +1.21\% & +6.29\% & +0.93\% & +1.46\% & +1.72\% \\ \bottomrule

\end{tabular}} \vspace{-1.5em}
\end{center}
\end{table*}

\vspace{-0.9em}
\section{Experiments}
\vspace{-0.5em}
\textbf{Baselines.} We compare Prompt-DDG with four categories of state-of-the-art methods. The first category is to directly extend those sequence-based approaches from single proteins to protein-protein interactions, including ESM-1v~\cite{meier2021language}, Position-Specific Scoring Matrix (PSSM), MSA Transformer~\cite{rao2021msa}, and Tranception~\cite{notin2022tranception}. The second category is the traditional energy-based approaches, including Rosetta~\cite{alford2017rosetta} Cartesian ddG and FoldX~\cite{delgado2019foldx}. The third category is supervised learning approaches, including DDGPred~\cite{shan2022deep} and a model that uses a self-attention-based network~\cite{jumper2021highly} as the encoder, but uses the MLP to directly predict $\Delta\Delta G$ (End-to-End). The fourth category is pre-training-based approaches, including ESM-1F~\cite{hsu2022learning}, two variants of MIF (MIF-$\Delta \text{logits}$ and MIF-Network)~\cite{yang2020graph}, two variants of RDE (RDE-Linear and RDE-Network)~\cite{luo2023rotamer}, DiffAffinity~\cite{liu2023predicting}, and a model that is pre-trained to predict the B-factor of residues and use predicted B-factors to predict $\Delta\Delta G$. A more detailed description of these methods and hyperparameter settings of our Prompt-DDG can be found in \textbf{Appendix F \& G}. 

\textbf{Datasets.} We evaluate the effectiveness of Prompt-DDG for $\Delta\Delta G$ prediction on the SKEMPI v2.0~\cite{jankauskaite2019skempi} dataset, the largest available annotated mutation dataset for protein complexes. The SKEMPI v2.0 dataset contains 7,085 amino acid mutations and corresponding changes in the thermodynamic parameters and kinetic rate constants, but it does not contain any structures of the mutated complexes. To avoid data leakage, we split the dataset into 3 folds by structure, each of which contains unique protein complexes. Then, we follow~\cite{luo2023rotamer} to perform 3-fold cross-validation to ensure that each data in SKEMPI2 is tested once. In terms of pre-training, ESM-1F is pre-trained on millions of predicted AF2 structures~\cite{jumper2021highly}, and the other six methods are all pre-trained on the PDB-REDO~\cite{joosten2014pdb_redo} dataset which contains 143k data. In contrast, Prompt-DDG directly learns prompts in a lightweight manner on the SKEMPI v2.0 dataset without requiring any additional pre-training data.

\vspace{-0.3em}
\textbf{Evaluation Metrics.} A total of seven metrics are used to comprehensively evaluate the performance of $\Delta\Delta G$ prediction, including five overall metrics: (1) Pearson correlation coefficient; (2) Spearman correlation coefficient; (3) Root Mean Squared Error (RMSE); (4) Mean Absolute Error (MAE); (5) AUROC. Since the correlation of specific protein complexes is often of greater interest in practice, we group the mutations by structure, calculate the Pearson and Spearman correlation coefficients for each structure separately, and report the average as two additional metrics.

\begin{table*}[!htbp]
\begin{center}
\vspace{-1em}
\caption{Performance comparison under single-, multi- and all-point mutation, where \textbf{bold} denotes the best metric under each setting.}
\vspace{0.3em}
\label{tab:2}
\resizebox{0.98\textwidth}{!}{
\begin{tabular}{lcl|cc|ccccc}

\toprule
\multirow{2}{*}{\textbf{Method}} & \multirow{2}{*}{\begin{tabular}[c]{@{}c@{}}\textbf{Pre-training}\\ \textbf{Dataset (Szie)}\end{tabular}} & \multirow{2}{*}{\textbf{Mutations}} & \multicolumn{2}{c}{\textbf{Per-Structure}} & \multicolumn{5}{c}{\textbf{Overall}} \\ \cmidrule(r){4-5} \cmidrule(r){6-10}
 &  &  & \textbf{Pearson} $\uparrow$ & \textbf{Spear.} $\uparrow$ & \textbf{Pearson} $\uparrow$ & \textbf{Spear.} $\uparrow$ & \textbf{RMSE} $\downarrow$ & \textbf{MAE} $\downarrow$ & \textbf{AUROC} $\uparrow$ \\ \midrule
\multirow{3}{*}{DDGPred} & \multirow{3}{*}{\XSolidBrush} & all & 0.3750 & 0.3407 & 0.6580 & 0.4687 & \textbf{1.4998} & 1.0821 & 0.6992 \\
 &  & single & 0.3711 & 0.3427 & 0.6515 & 0.4390 & 1.3285 & 0.9618 & 0.6858 \\
 &  & multiple & 0.3912 & 0.3896 & 0.5938 & 0.5150 & 2.1813 & 1.6699 & 0.7590 \\ \midrule
\multirow{3}{*}{End-to-End} & \multirow{3}{*}{\XSolidBrush} & all & 0.3873 & 0.3587 & 0.6373 & 0.4882 & 1.6198 & 1.1761 & 0.7172 \\
 &  & single & 0.3818 & 0.3426 & 0.6605 & 0.4594 & 1.3148 & 0.9569 & 0.7019 \\
 &  & multiple & 0.4178 & 0.4034 & 0.5858 & 0.4942 & 2.1971 & 1.7087 & 0.7532 \\ \midrule
\multirow{3}{*}{MIF-Network} & \multirow{3}{*}{\begin{tabular}[c]{@{}c@{}}PDB-REDO \\ (143k)\end{tabular}} & all & 0.3965 & 0.3509 & 0.6523 & 0.5134 & 1.5932 & 1.1469 & 0.7329 \\
 &  & single & 0.3952 & 0.3479 & 0.6667 & 0.4802 & 1.3052 & 0.9411 & 0.7175 \\
 &  & multiple & 0.3968 & 0.3789 & 0.6139 & 0.5370 & 2.1399 & 1.6422 & 0.7735 \\ \midrule
\multirow{3}{*}{RDE-Network} & \multirow{3}{*}{\begin{tabular}[c]{@{}c@{}}PDB-REDO \\ (143k)\end{tabular}} & all & 0.4448 & 0.4010 & 0.6447 & 0.5584 & 1.5799 & 1.1123 & 0.7454 \\
 &  & single & 0.4687 & 0.4333 & 0.6421 & 0.5271 & 1.3333 & 0.9392 & \textbf{0.7367} \\
 &  & multiple & 0.4233 & 0.3926 & 0.6288 & 0.5900 & 2.0980 & 1.5747 & 0.7749 \\ \midrule
\multirow{3}{*}{DiffAffinity} & \multirow{3}{*}{\begin{tabular}[c]{@{}c@{}}PDB-REDO \\ (143k)\end{tabular}} & all & 0.4220 & 0.3970 & 0.6690 & 0.5560 & 1.5350 & 1.0930 & 0.7440 \\
 &  & single & 0.4290 & 0.4090 & \textbf{0.6720} & 0.5230 & \textbf{1.2880} & 0.9230 & 0.7330 \\
 &  & multiple & 0.4140 & 0.3870 & 0.6500 & 0.6020 & 2.0510 & 1.5400 & 0.7840 \\ \midrule
\multicolumn{1}{c}{\multirow{3}{*}{Prompt-DDG}} & \multirow{3}{*}{\begin{tabular}[c]{@{}c@{}}SKEMPI v2.0 \\ (7k)\end{tabular}} & all & \textbf{0.4712} & \textbf{0.4257} & \textbf{0.6772} & \textbf{0.5910} & 1.5207 & \textbf{1.0770} & \textbf{0.7568} \\
\multicolumn{1}{c}{} &  & single & \textbf{0.4736} & \textbf{0.4392} & 0.6596 & \textbf{0.5450} & 1.3072 & \textbf{0.9191} & 0.7355 \\
\multicolumn{1}{c}{} &  & multiple & \textbf{0.4448} & \textbf{0.3961} & \textbf{0.6780} & \textbf{0.6433} & \textbf{1.9831} & \textbf{1.4837} & \textbf{0.8187} \\ \bottomrule

\end{tabular}} \vspace{-1em}
\end{center}
\end{table*}

\vspace{-0.5em}
\subsection{Comparison with State-of-The-Art Baselines}
\vspace{-0.5em}
We report the 7 evaluation metrics for the 15 methods on the SKEMPI v2.0 dataset in Table.~\ref{tab:1}, as well as the relative improvement of Prompt-GPT over the two leading methods, RDE-Network and DiffAffinity. It can be observed that (1) Prompt-DDG outperforms all baselines in 6 out of 7 evaluation metrics. In addition, it ranks second only in the RMSE metric, close to the state-of-the-art supervised method, DDGPred. (2) Despite not being pre-trained with any additional data, Prompt-DDG exceeds all pre-training-based methods across 7 metrics, which suggests that \textbf{specialized} microenvironmental prompts are more effective for $\Delta\Delta G$ prediction than \textbf{general} knowledge learned from protein pre-training. (3) Notably, Prompt-DDG achieves the most significant improvement on the two most critical metrics, the per-structure Pearson and Spearman correlations, demonstrating its greater potential for practical applications.

Furthermore, we select five superior methods from Table.~\ref{tab:1} based on a comprehensive consideration of the 7 metrics and compare Prompt-DDG with them under single-point, multi-point, and all-point mutations. The results reported in Table.~\ref{tab:2} show that Prompt-DDG achieves the best overall performance under the single-point mutation setting, ranking first in 4 out of 7 metrics. In practice, it is a common case to mutate multiple amino acids to reach the desired binding affinity, making the effect prediction of multi-point mutations very important. In particular, Prompt-DDG outperforms all other baselines, including RDE-Network and DiffAffinity, by a large margin in the multi-point mutation setting. The superiority of Prompt-GNN for multi-point mutation is twofold: (1) it generates prompts for the microenvironment around each mutation separately, which captures more fine-grained local (rather than global) differences; and (2) the conformation of a complex with multiple mutations is more variable than that with a single mutation, and Prompt-DDG is good at modeling the effects of each mutation on its local microenvironmental conformation.

\vspace{-0.8em}
\subsection{Visualization for Correlation Analysis}
\begin{figure*}[!tbp]
	\begin{center}
        \includegraphics[width=0.24\linewidth]{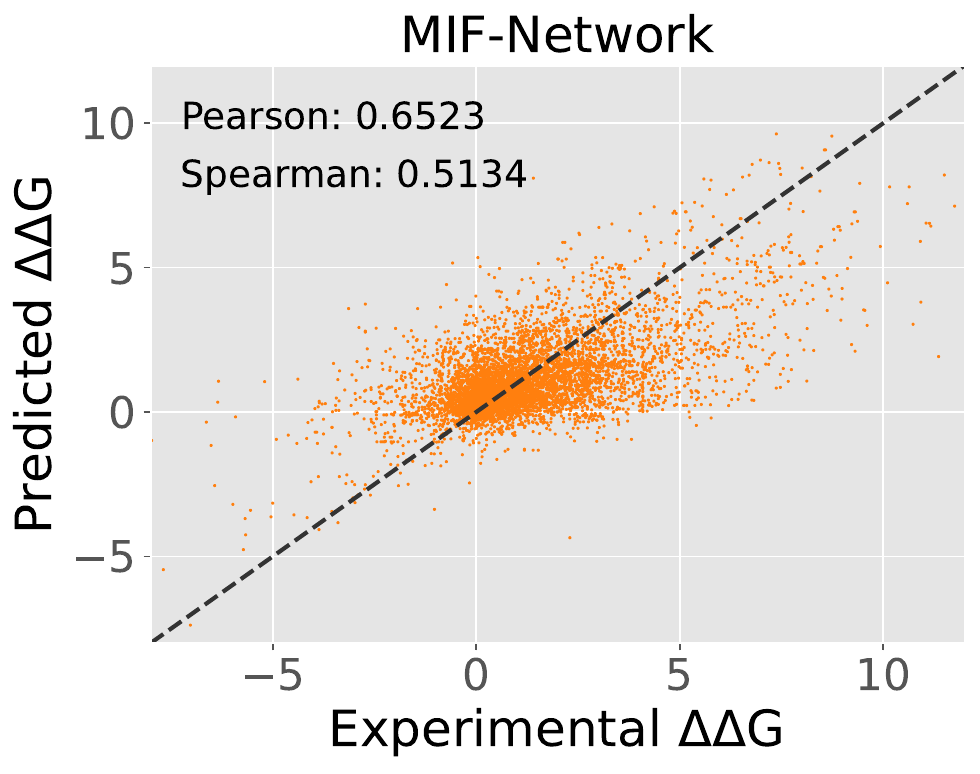}
        \includegraphics[width=0.24\linewidth]{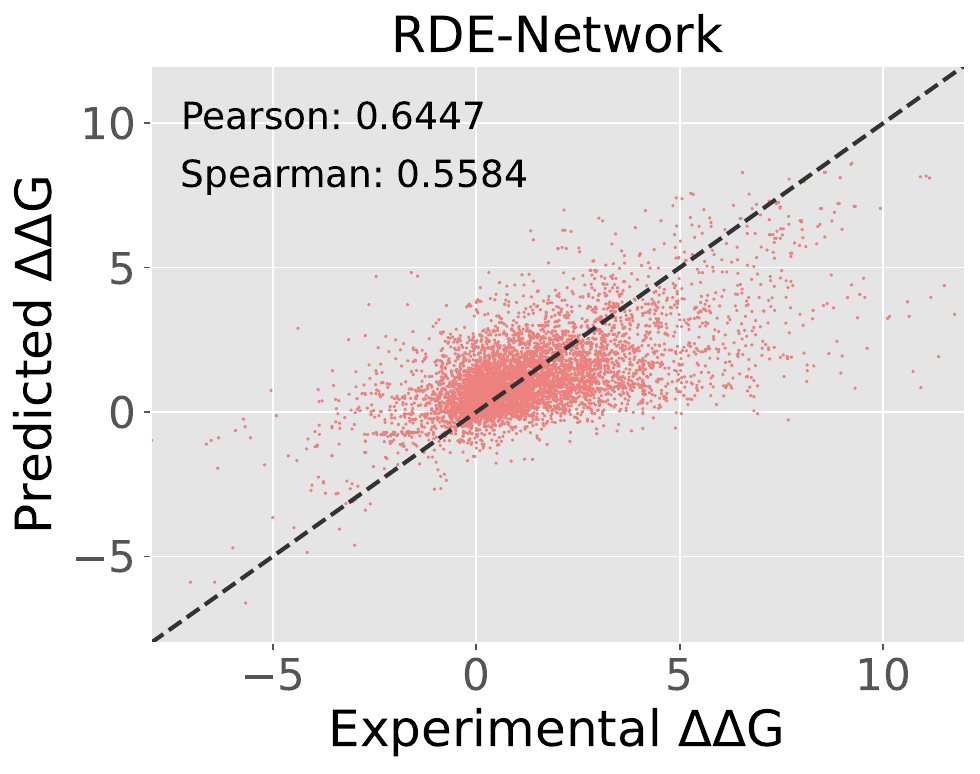}
        \includegraphics[width=0.24\linewidth]{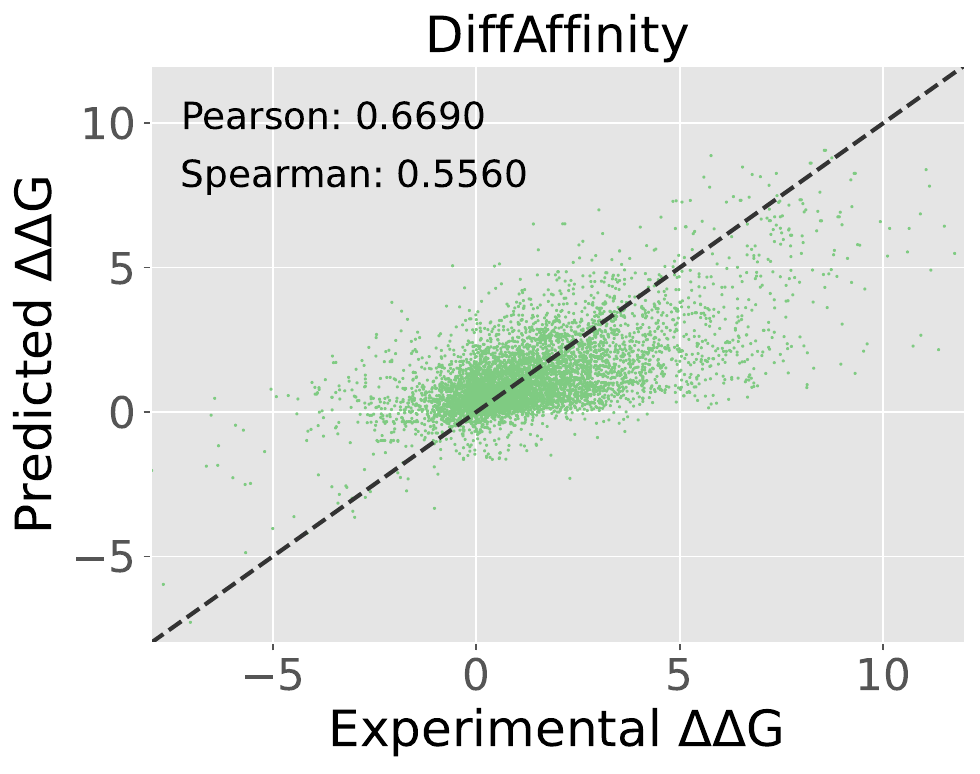}
        \includegraphics[width=0.24\linewidth]{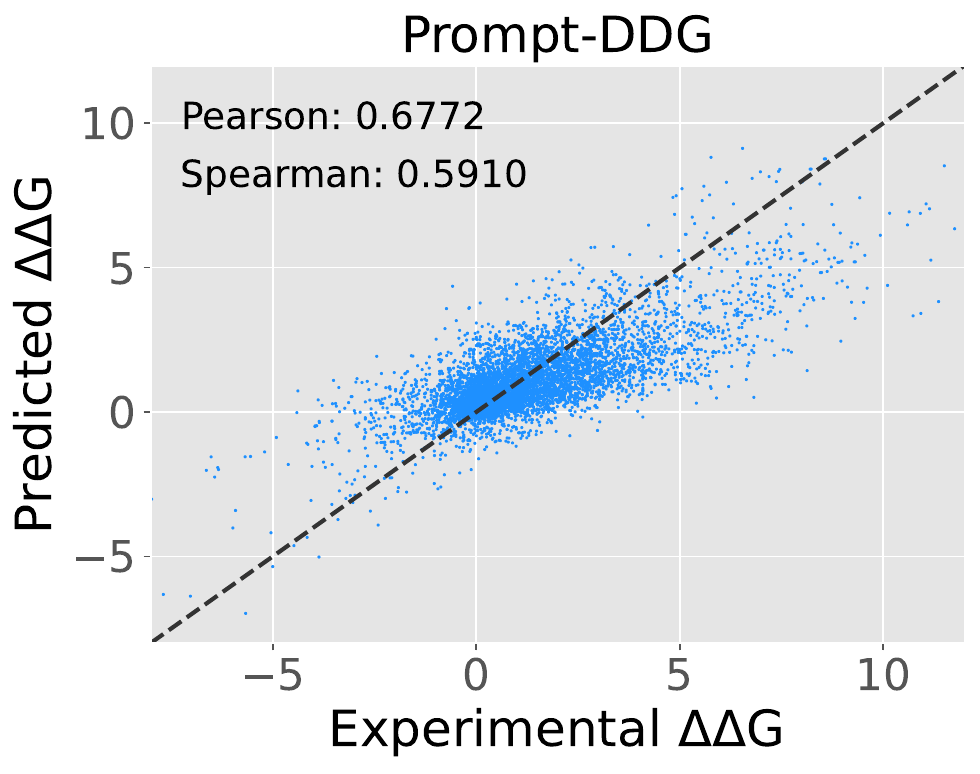}
	\end{center}
    \vspace{-1em}
	\caption{A comparison of correlations between experimental $\Delta\Delta G$ and $\Delta\Delta G$ predicted by four representative methods.}
    \vspace{-1em}
	\label{fig:4}
\end{figure*}

\begin{figure*}[!tbp]
	\begin{center}
        \includegraphics[width=0.4\linewidth]{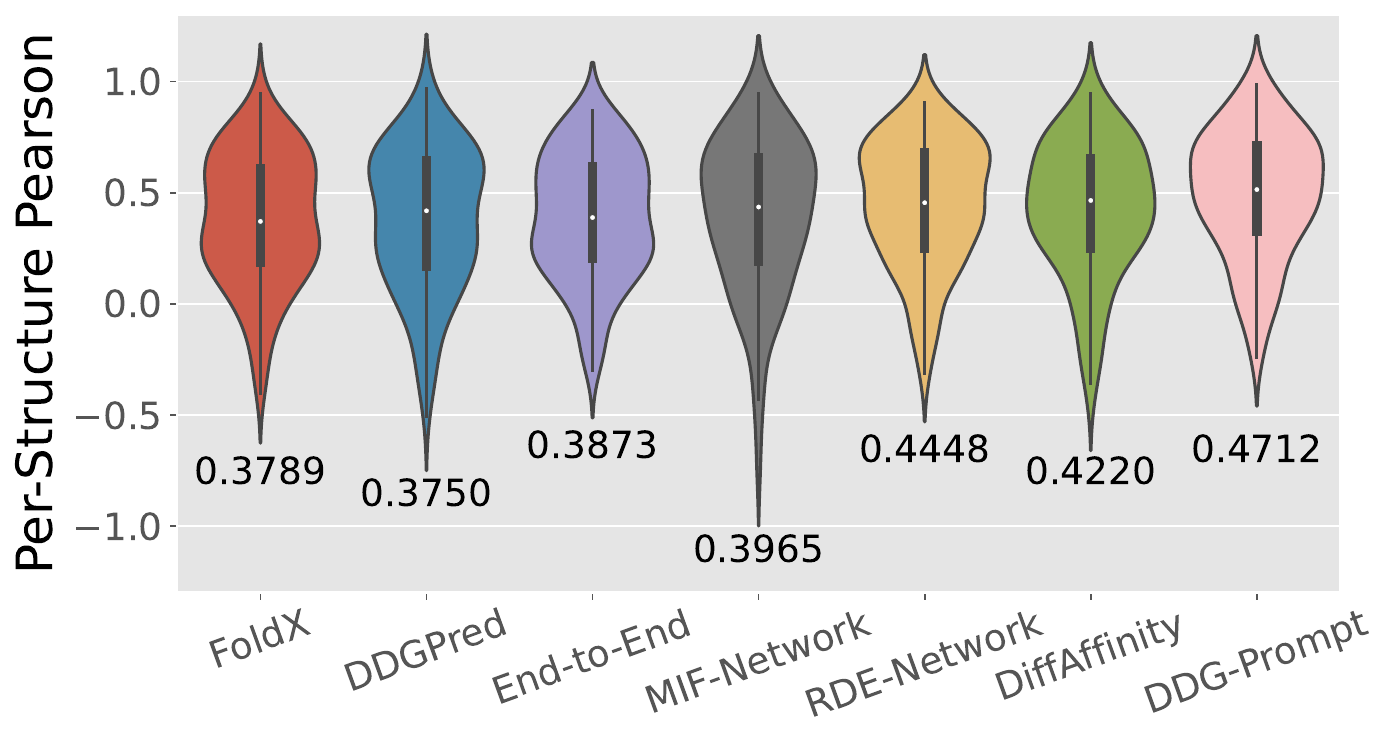}
        \includegraphics[width=0.4\linewidth]{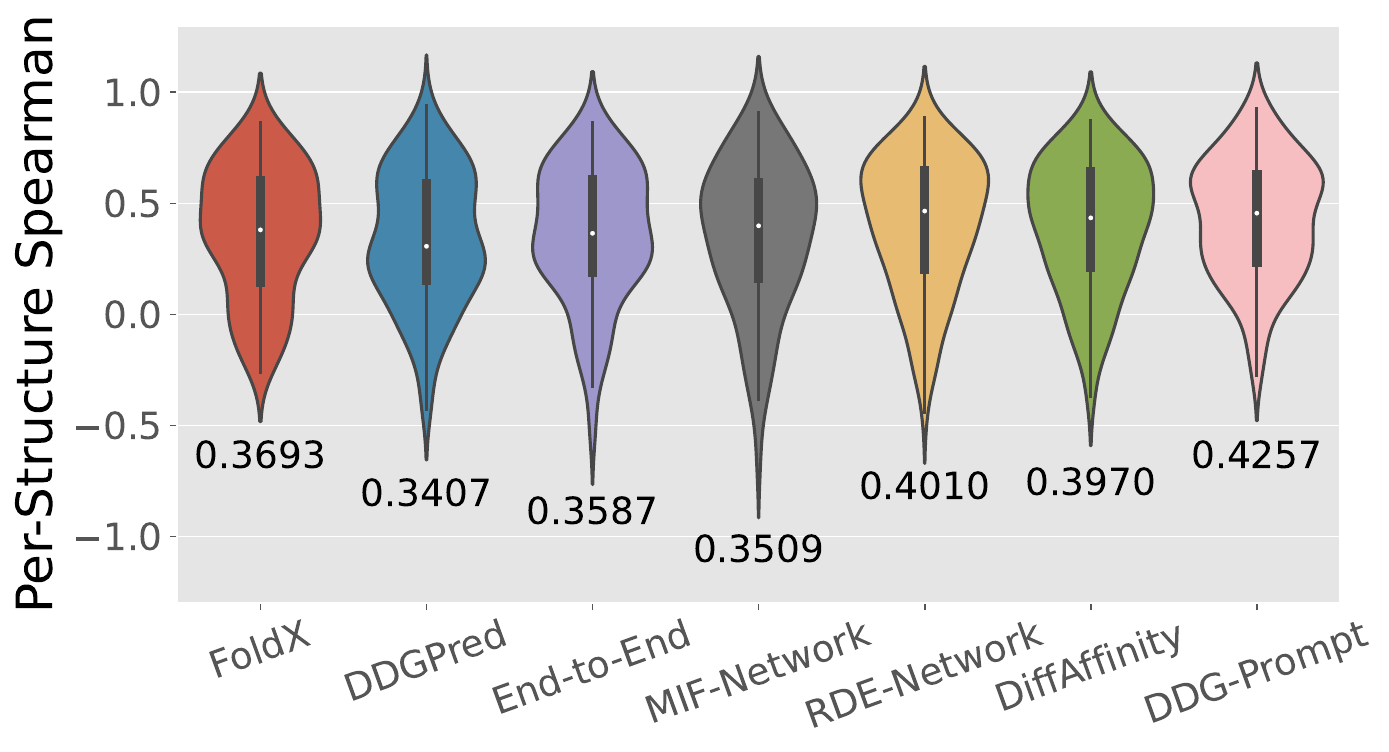}
	\end{center}
    \vspace{-1.5em}
	\caption{Distributions of per-structure Pearson correlation scores and Spearman correlation scores for seven representative methods.}
	\label{fig:5}
\end{figure*}

\vspace{-0.5em}
We present in Figure.~\ref{fig:4} the scatter plots of experimental and predicted $\Delta\Delta G$ for four representative methods, MIF-Network, RDE-Network, DiffAffinity, and Prompt-DDG, as well as their overall Pearson and Spearman correlation scores. It can be seen that Prompt-DDG performs better than the other three methods, both for qualitative visualization and quantitative metrics. Moreover, we provide the distribution of per-structure Pearson and Spearman correlation scores in Figure.~\ref{fig:5}, as well as the average results across all structures. We find that Prompt-DDG not only has the best average performance, but also that its distribution is mostly centered on high correlations and has fewer low-correlation structures. Due to space limitations, a comparison of Prompt-DDG's visualizations for single-point, multi-point, and all-point mutations is available in \textbf{Appendix H}.

\begin{table*}[!htbp]
\begin{center}
\vspace{-1em}
\caption{Ablation study on different types of microenvironmental prompts, where \textbf{bold} and \underline{underline} denote the best and second metrics.}
\vspace{0.3em}
\label{tab:3}
\resizebox{0.9\textwidth}{!}{
\begin{tabular}{lccc|cc|ccccc}

\toprule
\multirow{2}{*}{\textbf{Method}} & \multicolumn{3}{c}{\textbf{Prompt}} & \multicolumn{2}{c}{\textbf{Per-Structure}} & \multicolumn{5}{c}{\textbf{Overall}} \\ \cmidrule(r){2-4} \cmidrule(r){5-6} \cmidrule(r){7-11}
 & 1D Type & 2D Angle & 3D Coor. & \textbf{Pearson} $\uparrow$ & \textbf{Spear.} $\uparrow$ & \textbf{Pearson} $\uparrow$ & \textbf{Spear.} $\uparrow$ & \textbf{RMSE} $\downarrow$ & \textbf{MAE} $\downarrow$ & \textbf{AUROC} $\uparrow$ \\ \midrule
w/o Prompt & \XSolidBrush & \XSolidBrush & \XSolidBrush & 0.4114 & 0.3685 & 0.6494 & 0.5417 & 1.5716 & 1.1263 & 0.7309 \\ \midrule
\multirow{3}{*}{Single Prompt} & \Checkmark & \XSolidBrush & \XSolidBrush & 0.4263 & 0.3784 & 0.6436 & 0.5397 & 1.5518 & 1.1214 & 0.7352 \\
 & \XSolidBrush & \Checkmark & \XSolidBrush & 0.4462 & 0.4013 & 0.6642 & 0.5674 & 1.5450 & 1.1150 & 0.7456 \\
 & \XSolidBrush & \XSolidBrush & \Checkmark & 0.4583 & 0.4129 & \underline{0.6696} & 0.5745 & \underline{1.5350} & 1.1054 & 0.7532 \\ \midrule
Average & \Checkmark & \Checkmark & \Checkmark & \underline{0.4652} & \underline{0.4148} & 0.6673 & \underline{0.5791} & 1.5394 & \underline{1.0874} & \textbf{0.7587} \\
Weighted & \Checkmark & \Checkmark & \Checkmark & \textbf{0.4712} & \textbf{0.4257} & \textbf{0.6772} & \textbf{0.5910} & \textbf{1.5207} & \textbf{1.0770} & \underline{0.7568} \\ \bottomrule

\end{tabular}} \vspace{-1em}
\end{center}
\end{table*}

\vspace{-0.8em}
\subsection{Ablation Study and Hyperparametric Sensitivity}
\vspace{-0.5em}
\textbf{Ablation Study.} We conduct a comprehensive evaluation on the necessity of prompts, the importance of different (structural scales) prompts, and the prompt combination schemes. From the results reported in Table.~\ref{tab:3}, three important observations can be made: (1) Three kinds of prompts characterize different structural scales of the microenvironment, each playing a different role and helpful in improving performance compared to the ones without any prompts. (2) Compared to the residue types and angular statistics of the microenvironment, how the mutation affects the local conformation is more important, and thus the corresponding prompt brings a huger performance gain than the other two prompts. (3) Combining all three kinds of prompts, either by simple averaging or weighted adaptation, outperforms any single kind of prompt. Besides, the weighted prompt adaptation module proposed in this paper, albeit lightweight, outperforms the average combination by a wide margin.

\textbf{Hyperparametric Sensitivity.} We studied the sensitivity of Prompt-DDG to two hyperparameters, codebook size $|\mathcal{A}|$ and mask ratio $r$, in Table.~\ref{tab:4} and \ref{tab:5}. A small codebook size, e.g., $|\mathcal{A}|=64$, leads to sub-optimal performance due to the incapacity to cover the diversity of the microenvironment. Conversely, setting $|\mathcal{A}|$ too large may result in redundant codebooks and high computational cost. In addition, we find that the removal of the microenvironment masking, i.e., setting the mask ratio to 0.0, leads to a sharp performance drop. In practice, a mask ratio of 0.1 or 0.2 usually yields good performance, since a small mask ratio, e.g. 0.05, weakens the contribution of masked modeling, while a large mask ratio, e.g. 0.3, hinders the prompt codebook learning.

\begin{table}[!tbp]
\begin{center}
\vspace{-1em}
\caption{Hyperparameter sensitivity analysis on codebook size $|\mathcal{A}|$.}
\vspace{0.5em}
\label{tab:4}
\resizebox{0.9\columnwidth}{!}{
\begin{tabular}{c|cccc}
\toprule
\multirow{2}{*}{\begin{tabular}[c]{@{}l@{}}Codebook \\ Size $|\mathcal{A}|$\end{tabular}} & \multicolumn{2}{c}{\textbf{Per-Structure}} & \multicolumn{2}{c}{\textbf{Overall}} \\ \cmidrule(r){2-3} \cmidrule(r){4-5} 
 & \textbf{Pearson} $\uparrow$ & \textbf{Spear.} $\uparrow$ & \textbf{Pearson} $\uparrow$ & \textbf{Spear.} $\uparrow$ \\ \midrule
64 & 0.4219 & 0.3887 & 0.6453 & 0.5512 \\
128 & \underline{0.4575} & \underline{0.4132} & \underline{0.6735} & \textbf{0.5971} \\
256 & \textbf{0.4712} & \textbf{0.4257} & \textbf{0.6772} & \underline{0.5910} \\
512 & 0.4457 & 0.4077 & 0.6587 & 0.5772 \\
1024 & 0.4289 & 0.3916 & 0.6557 & 0.5691 \\ \bottomrule

\end{tabular}} \vspace{-1.5em}
\end{center}
\end{table} 
\begin{table}[!tbp]
\begin{center}
\vspace{-1em}
\caption{Hyperparameter sensitivity analysis on mask ratio $r$.}
\vspace{0.5em}
\label{tab:5}
\resizebox{0.9\columnwidth}{!}{
\begin{tabular}{c|cccc}
\toprule
\multirow{2}{*}{Mask Ratio} & \multicolumn{2}{c}{\textbf{Per-Structure}} & \multicolumn{2}{c}{\textbf{Overall}} \\ \cmidrule(r){2-3} \cmidrule(r){4-5} 
 & \textbf{Pearson} $\uparrow$ & \textbf{Spear.} $\uparrow$ & \textbf{Pearson} $\uparrow$ & \textbf{Spear.} $\uparrow$ \\ \midrule
0.00 & 0.4462 & 0.4018 & 0.6467 & 0.5610 \\
0.05 & 0.4652 & 0.4186 & 0.6661 & 0.5795 \\
0.10 & \underline{0.4712} & \underline{0.4257} & \textbf{0.6772} & \textbf{0.5910} \\
0.20 & \textbf{0.4738} & \textbf{0.4358} & \underline{0.6743} & \underline{0.5833} \\
0.30 & 0.4584 & 0.4136 & 0.6593 & 0.5683 \\ \bottomrule

\end{tabular}} \vspace{-1em}
\end{center}
\end{table}

\vspace{-0.5em}
\subsection{Antibody Optimization against SARS-CoV-2}
\vspace{-0.5em}
An important usage scenario for $\Delta\Delta G$ prediction is to identify those desirable mutations, usually with high binding affinity or neutralization, from a pool of potential mutations. In this subsection, we take the optimization of human antibodies against SARS-CoV-2 as a case study. We predict $\Delta\Delta G$s for 494 possible single-point mutations in the 26 sites within the CDR region of the antibody heavy chain, and rank them in ascending order (lowest $\Delta\Delta G$ in the top). Then, we report in Table.~\ref{tab:6} the ranking of five favorable mutations that have been previously shown to help enhance neutralization~\cite{shan2022deep}. The results in Table.~\ref{tab:6} show that only Prompt-DDG can successfully identify the four important mutations with rankings smaller than 10\% (in \textbf{bold}). Moreover, Prompt-DDG achieves the highest average ranking, 7.57\% and 13.80\% higher than RDE-Network and DiffAffinity, respectively. More importantly, only Prompt-DDG ranks all five favorable mutations in the top 40\%, suggesting good generalizability to different antibodies.

\begin{table}[!tbp]
\begin{center}
\vspace{-0.5em}
\caption{Rankings of the five favorable mutations on the antibody against SARS-CoV-2 by various $\Delta\Delta G$ prediction methods.}
\vspace{-0.5em}
\label{tab:6}
\resizebox{\columnwidth}{!}{
\begin{tabular}{l|ccccc|c}
\midrule
\textbf{Method} & TH31W & AH53F & NH57L & RH103M & LH104F & Average \\ \midrule
Rosetta & 10.73\% & 76.72\% & 93.93\% & 11.34\% & 27.94\% & 44.13\% \\
FoldX & 13.56\% & \textbf{6.88\%} & \textbf{5.67\%} & 16.60\% & 66.19\% & 21.78\% \\
DDGPred & 68.22\% & \textbf{2.63\%} & 12.35\% & \textbf{8.30\%} & \textbf{8.50\%} & 20.00\% \\
End-to-End & 29.96\% & \textbf{2.02\%} & 14.17\% & 52.43\% & 17.21\% & 23.16\% \\ \midrule
MIF-Net. & 24.49\% & \textbf{4.05\%} & \textbf{6.48\%} & 80.36\% & 36.23\% & 30.32\% \\
RDE-Net. & \textbf{1.62\%} & \textbf{2.02\%} & 20.65\% & 61.54\% & \textbf{5.47\%} & 18.26\% \\
DiffAffinity & \textbf{7.28\%} & \textbf{3.64\%} & 18.82\% & 81.78\% & 10.93\% & 24.49\% \\ \midrule
Prompt-DDG & \textbf{2.02\%} & \textbf{6.88\%} & \textbf{3.24\%} & 34.81\% & \textbf{6.48\%} & 10.69\% \\ \bottomrule

\end{tabular}} \vspace{-1em}
\end{center}
\end{table}

\vspace{-0.5em}
\section{Conclusion}
\vspace{-0.3em}
In this paper, we propose a novel Prompt-DDG framework for efficient and effective $\Delta\Delta G$ prediction. Specifically, a hierarchical prompt codebook is constructed and pre-trained by masked microenvironment modeling to cover the different structural scales of the microenvironment around each mutation. The microenvironment-aware prompts generated for each mutation flexibly provide wild-type and mutated complexes with multi-scale structural information about their microenvironmental differences. Extensive experiments have shown that Prompt-DDG achieves superior performance and efficiency over existing methods in terms of both mutation effect prediction and antibody optimization.

\clearpage
\section*{Acknowledgments}
This work is supported by National Science and Technology Major Project (No. 2022ZD0115101), National Natural Science Foundation of China Project (No. U21A20427), and Project (No. WU2022A009) from Center of Synthetic Biology and Integrated Bioengineering of Westlake University.

\section*{Code Resources}
Codes of this work are publicly available at: \url{https://github.com/LirongWu/Prompt-DDG}.

\section*{Impact Statement}
This paper presents work whose goal is to advance the field of machine learning for protein modeling. Prompt-DDG is expected to play significant roles in many potential biological applications. Firstly, Prompt-DDG, as a general DDG predictor trained on lots of antigen-antibody complexes with different sequences and structures, can be used as a computationally validated means for virtual screening of candidate antibodies, which helps to increase the success rate of antibody design and reduce the costs. Secondly, Prompt-DDG can be used as an explainer for discovering key functional sites on antibodies. This is because Prompt-DDG works by comparing the microenvironmental differences around each mutation, from which it locates salient mutations that are critical for binding energy. These salient mutations may be potential functional sites. Finally, Prompt-DDG can be used as a data augmentor to address data scarcity in antibody design/optimization by generating high-quality antibody data. Despite the great successes, limitations still exist. As a manuscript submitted to the ML venue, we did not spend much space discussing biology-related details, especially those about wet experiments. Moreover, Prompt-DDG mainly focuses on the interaction between two proteins and is limited in the prediction of mutational effects for multimeric proteins or substrate-related enzymes.

\bibliography{example_paper}

\begin{thebibliography}{41}
\providecommand{\natexlab}[1]{#1}
\providecommand{\url}[1]{\texttt{#1}}
\expandafter\ifx\csname urlstyle\endcsname\relax
  \providecommand{\doi}[1]{doi: #1}\else
  \providecommand{\doi}{doi: \begingroup \urlstyle{rm}\Url}\fi

\bibitem[Alford et~al.(2017)Alford, Leaver-Fay, Jeliazkov, O’Meara, DiMaio, Park, Shapovalov, Renfrew, Mulligan, Kappel, et~al.]{alford2017rosetta}
Alford, R.~F., Leaver-Fay, A., Jeliazkov, J.~R., O’Meara, M.~J., DiMaio, F.~P., Park, H., Shapovalov, M.~V., Renfrew, P.~D., Mulligan, V.~K., Kappel, K., et~al.
\newblock The rosetta all-atom energy function for macromolecular modeling and design.
\newblock \emph{Journal of chemical theory and computation}, 13\penalty0 (6):\penalty0 3031--3048, 2017.

\bibitem[Bengio et~al.(2013)Bengio, L{\'e}onard, and Courville]{bengio2013estimating}
Bengio, Y., L{\'e}onard, N., and Courville, A.
\newblock Estimating or propagating gradients through stochastic neurons for conditional computation.
\newblock \emph{arXiv preprint arXiv:1308.3432}, 2013.

\bibitem[Delgado et~al.(2019)Delgado, Radusky, Cianferoni, and Serrano]{delgado2019foldx}
Delgado, J., Radusky, L.~G., Cianferoni, D., and Serrano, L.
\newblock Foldx 5.0: working with rna, small molecules and a new graphical interface.
\newblock \emph{Bioinformatics}, 35\penalty0 (20):\penalty0 4168--4169, 2019.

\bibitem[Elfwing et~al.(2018)Elfwing, Uchibe, and Doya]{elfwing2018sigmoid}
Elfwing, S., Uchibe, E., and Doya, K.
\newblock Sigmoid-weighted linear units for neural network function approximation in reinforcement learning.
\newblock \emph{Neural networks}, 107:\penalty0 3--11, 2018.

\bibitem[Frazer et~al.(2021)Frazer, Notin, Dias, Gomez, Min, Brock, Gal, and Marks]{frazer2021disease}
Frazer, J., Notin, P., Dias, M., Gomez, A., Min, J.~K., Brock, K., Gal, Y., and Marks, D.~S.
\newblock Disease variant prediction with deep generative models of evolutionary data.
\newblock \emph{Nature}, 599\penalty0 (7883):\penalty0 91--95, 2021.

\bibitem[Gao et~al.(2024)Gao, Tan, Zhang, Chen, Wu, and Li]{gao2024proteininvbench}
Gao, Z., Tan, C., Zhang, Y., Chen, X., Wu, L., and Li, S.~Z.
\newblock Proteininvbench: Benchmarking protein inverse folding on diverse tasks, models, and metrics.
\newblock \emph{Advances in Neural Information Processing Systems}, 36, 2024.

\bibitem[Geng et~al.(2019)Geng, Vangone, Folkers, Xue, and Bonvin]{geng2019isee}
Geng, C., Vangone, A., Folkers, G.~E., Xue, L.~C., and Bonvin, A.~M.
\newblock isee: Interface structure, evolution, and energy-based machine learning predictor of binding affinity changes upon mutations.
\newblock \emph{Proteins: Structure, Function, and Bioinformatics}, 87\penalty0 (2):\penalty0 110--119, 2019.

\bibitem[Hsu et~al.(2022)Hsu, Verkuil, Liu, Lin, Hie, Sercu, Lerer, and Rives]{hsu2022learning}
Hsu, C., Verkuil, R., Liu, J., Lin, Z., Hie, B., Sercu, T., Lerer, A., and Rives, A.
\newblock Learning inverse folding from millions of predicted structures.
\newblock In \emph{International Conference on Machine Learning}, pp.\  8946--8970. PMLR, 2022.

\bibitem[Hu et~al.(2021)Hu, Wang, Huang, Hu, and You]{hu2021survey}
Hu, L., Wang, X., Huang, Y.-A., Hu, P., and You, Z.-H.
\newblock A survey on computational models for predicting protein--protein interactions.
\newblock \emph{Briefings in bioinformatics}, 22\penalty0 (5):\penalty0 bbab036, 2021.

\bibitem[Huang et~al.(2023)Huang, Wu, Lin, Zheng, Wang, and Li]{huang2023data}
Huang, Y., Wu, L., Lin, H., Zheng, J., Wang, G., and Li, S.~Z.
\newblock Data-efficient protein 3d geometric pretraining via refinement of diffused protein structure decoy.
\newblock \emph{arXiv preprint arXiv:2302.10888}, 2023.

\bibitem[Huang et~al.(2024)Huang, Li, Wu, Su, Lin, Zhang, Liu, Gao, Zheng, and Li]{huang2024protein}
Huang, Y., Li, S., Wu, L., Su, J., Lin, H., Zhang, O., Liu, Z., Gao, Z., Zheng, J., and Li, S.~Z.
\newblock Protein 3d graph structure learning for robust structure-based protein property prediction.
\newblock In \emph{Proceedings of the AAAI Conference on Artificial Intelligence}, volume~38, pp.\  12662--12670, 2024.

\bibitem[Huber(1992)]{huber1992robust}
Huber, P.~J.
\newblock Robust estimation of a location parameter.
\newblock In \emph{Breakthroughs in statistics: Methodology and distribution}, pp.\  492--518. Springer, 1992.

\bibitem[Jankauskait{\.e} et~al.(2019)Jankauskait{\.e}, Jim{\'e}nez-Garc{\'\i}a, Dapk{\=u}nas, Fern{\'a}ndez-Recio, and Moal]{jankauskaite2019skempi}
Jankauskait{\.e}, J., Jim{\'e}nez-Garc{\'\i}a, B., Dapk{\=u}nas, J., Fern{\'a}ndez-Recio, J., and Moal, I.~H.
\newblock Skempi 2.0: an updated benchmark of changes in protein--protein binding energy, kinetics and thermodynamics upon mutation.
\newblock \emph{Bioinformatics}, 35\penalty0 (3):\penalty0 462--469, 2019.

\bibitem[Joosten et~al.(2014)Joosten, Long, Murshudov, and Perrakis]{joosten2014pdb_redo}
Joosten, R.~P., Long, F., Murshudov, G.~N., and Perrakis, A.
\newblock The pdb\_redo server for macromolecular structure model optimization.
\newblock \emph{IUCrJ}, 1\penalty0 (4):\penalty0 213--220, 2014.

\bibitem[Jumper et~al.(2021)Jumper, Evans, Pritzel, Green, Figurnov, Ronneberger, Tunyasuvunakool, Bates, {\v{Z}}{\'\i}dek, Potapenko, et~al.]{jumper2021highly}
Jumper, J., Evans, R., Pritzel, A., Green, T., Figurnov, M., Ronneberger, O., Tunyasuvunakool, K., Bates, R., {\v{Z}}{\'\i}dek, A., Potapenko, A., et~al.
\newblock Highly accurate protein structure prediction with alphafold.
\newblock \emph{Nature}, 596\penalty0 (7873):\penalty0 583--589, 2021.

\bibitem[Kastritis \& Bonvin(2013)Kastritis and Bonvin]{kastritis2013binding}
Kastritis, P.~L. and Bonvin, A.~M.
\newblock On the binding affinity of macromolecular interactions: daring to ask why proteins interact.
\newblock \emph{Journal of The Royal Society Interface}, 10\penalty0 (79):\penalty0 20120835, 2013.

\bibitem[Kingma \& Ba(2014)Kingma and Ba]{kingma2014adam}
Kingma, D.~P. and Ba, J.
\newblock Adam: A method for stochastic optimization.
\newblock \emph{arXiv preprint arXiv:1412.6980}, 2014.

\bibitem[Kong et~al.(2022)Kong, Huang, and Liu]{kong2022conditional}
Kong, X., Huang, W., and Liu, Y.
\newblock Conditional antibody design as 3d equivariant graph translation.
\newblock \emph{arXiv preprint arXiv:2208.06073}, 2022.

\bibitem[Lei et~al.(2023)Lei, Garcia, Tan, Teo, Wang, Zhang, Luo, Nair, Peng, and Wu]{lei2023mutational}
Lei, R., Garcia, A.~H., Tan, T.~J., Teo, Q.~W., Wang, Y., Zhang, X., Luo, S., Nair, S.~K., Peng, J., and Wu, N.~C.
\newblock Mutational fitness landscape of human influenza h3n2 neuraminidase.
\newblock \emph{Cell reports}, 42\penalty0 (1), 2023.

\bibitem[Li et~al.(2016)Li, Simonetti, Goncearenco, and Panchenko]{li2016mutabind}
Li, M., Simonetti, F.~L., Goncearenco, A., and Panchenko, A.~R.
\newblock Mutabind estimates and interprets the effects of sequence variants on protein--protein interactions.
\newblock \emph{Nucleic acids research}, 44\penalty0 (W1):\penalty0 W494--W501, 2016.

\bibitem[Lin et~al.(2024)Lin, Huang, Zhang, Liu, Wu, Li, Chen, and Li]{lin2024functional}
Lin, H., Huang, Y., Zhang, O., Liu, Y., Wu, L., Li, S., Chen, Z., and Li, S.~Z.
\newblock Functional-group-based diffusion for pocket-specific molecule generation and elaboration.
\newblock \emph{Advances in Neural Information Processing Systems}, 36, 2024.

\bibitem[Lin et~al.(2023)Lin, Akin, Rao, Hie, Zhu, Lu, Smetanin, Verkuil, Kabeli, Shmueli, et~al.]{lin2023evolutionary}
Lin, Z., Akin, H., Rao, R., Hie, B., Zhu, Z., Lu, W., Smetanin, N., Verkuil, R., Kabeli, O., Shmueli, Y., et~al.
\newblock Evolutionary-scale prediction of atomic-level protein structure with a language model.
\newblock \emph{Science}, 379\penalty0 (6637):\penalty0 1123--1130, 2023.

\bibitem[Liu et~al.(2023)Liu, Zhu, Ren, Yu, Bu, and Zhang]{liu2023predicting}
Liu, S., Zhu, T., Ren, M., Yu, C., Bu, D., and Zhang, H.
\newblock Predicting mutational effects on protein-protein binding via a side-chain diffusion probabilistic model.
\newblock \emph{arXiv preprint arXiv:2310.19849}, 2023.

\bibitem[Lu et~al.(2020)Lu, Zhou, He, Jiang, Peng, Tong, and Shi]{lu2020recent}
Lu, H., Zhou, Q., He, J., Jiang, Z., Peng, C., Tong, R., and Shi, J.
\newblock Recent advances in the development of protein--protein interactions modulators: mechanisms and clinical trials.
\newblock \emph{Signal transduction and targeted therapy}, 5\penalty0 (1):\penalty0 213, 2020.

\bibitem[Luo et~al.(2023)Luo, Su, Wu, Su, Peng, and Ma]{luo2023rotamer}
Luo, S., Su, Y., Wu, Z., Su, C., Peng, J., and Ma, J.
\newblock Rotamer density estimator is an unsupervised learner of the effect of mutations on protein-protein interaction.
\newblock \emph{bioRxiv}, pp.\  2023--02, 2023.

\bibitem[Luo et~al.(2021)Luo, Jiang, Yu, Liu, Vo, Ding, Su, Qian, Zhao, and Peng]{luo2021ecnet}
Luo, Y., Jiang, G., Yu, T., Liu, Y., Vo, L., Ding, H., Su, Y., Qian, W.~W., Zhao, H., and Peng, J.
\newblock Ecnet is an evolutionary context-integrated deep learning framework for protein engineering.
\newblock \emph{Nature communications}, 12\penalty0 (1):\penalty0 5743, 2021.

\bibitem[Marchand et~al.(2022)Marchand, Van Hall-Beauvais, and Correia]{marchand2022computational}
Marchand, A., Van Hall-Beauvais, A.~K., and Correia, B.~E.
\newblock Computational design of novel protein--protein interactions--an overview on methodological approaches and applications.
\newblock \emph{Current Opinion in Structural Biology}, 74:\penalty0 102370, 2022.

\bibitem[Meier et~al.(2021)Meier, Rao, Verkuil, Liu, Sercu, and Rives]{meier2021language}
Meier, J., Rao, R., Verkuil, R., Liu, J., Sercu, T., and Rives, A.
\newblock Language models enable zero-shot prediction of the effects of mutations on protein function.
\newblock \emph{Advances in Neural Information Processing Systems}, 34:\penalty0 29287--29303, 2021.

\bibitem[Murphy \& Weaver(2016)Murphy and Weaver]{murphy2016janeway}
Murphy, K. and Weaver, C.
\newblock \emph{Janeway's immunobiology}.
\newblock Garland science, 2016.

\bibitem[Notin et~al.(2022)Notin, Dias, Frazer, Hurtado, Gomez, Marks, and Gal]{notin2022tranception}
Notin, P., Dias, M., Frazer, J., Hurtado, J.~M., Gomez, A.~N., Marks, D., and Gal, Y.
\newblock Tranception: protein fitness prediction with autoregressive transformers and inference-time retrieval.
\newblock In \emph{International Conference on Machine Learning}, pp.\  16990--17017. PMLR, 2022.

\bibitem[Park et~al.(2016)Park, Bradley, Greisen~Jr, Liu, Mulligan, Kim, Baker, and DiMaio]{park2016simultaneous}
Park, H., Bradley, P., Greisen~Jr, P., Liu, Y., Mulligan, V.~K., Kim, D.~E., Baker, D., and DiMaio, F.
\newblock Simultaneous optimization of biomolecular energy functions on features from small molecules and macromolecules.
\newblock \emph{Journal of chemical theory and computation}, 12\penalty0 (12):\penalty0 6201--6212, 2016.

\bibitem[Rao et~al.(2021)Rao, Liu, Verkuil, Meier, Canny, Abbeel, Sercu, and Rives]{rao2021msa}
Rao, R.~M., Liu, J., Verkuil, R., Meier, J., Canny, J., Abbeel, P., Sercu, T., and Rives, A.
\newblock Msa transformer.
\newblock In \emph{International Conference on Machine Learning}, pp.\  8844--8856. PMLR, 2021.

\bibitem[Schymkowitz et~al.(2005)Schymkowitz, Borg, Stricher, Nys, Rousseau, and Serrano]{schymkowitz2005foldx}
Schymkowitz, J., Borg, J., Stricher, F., Nys, R., Rousseau, F., and Serrano, L.
\newblock The foldx web server: an online force field.
\newblock \emph{Nucleic acids research}, 33\penalty0 (suppl\_2):\penalty0 W382--W388, 2005.

\bibitem[Shan et~al.(2022)Shan, Luo, Yang, Hong, Su, Ding, Fu, Li, Chen, Ma, et~al.]{shan2022deep}
Shan, S., Luo, S., Yang, Z., Hong, J., Su, Y., Ding, F., Fu, L., Li, C., Chen, P., Ma, J., et~al.
\newblock Deep learning guided optimization of human antibody against sars-cov-2 variants with broad neutralization.
\newblock \emph{Proceedings of the National Academy of Sciences}, 119\penalty0 (11):\penalty0 e2122954119, 2022.

\bibitem[Tan et~al.(2024)Tan, Gao, Wu, Xia, Zheng, Yang, Liu, Hu, and Li]{tan2024cross}
Tan, C., Gao, Z., Wu, L., Xia, J., Zheng, J., Yang, X., Liu, Y., Hu, B., and Li, S.~Z.
\newblock Cross-gate mlp with protein complex invariant embedding is a one-shot antibody designer.
\newblock In \emph{Proceedings of the AAAI Conference on Artificial Intelligence}, volume~38, pp.\  15222--15230, 2024.

\bibitem[Van Den~Oord et~al.(2017)Van Den~Oord, Vinyals, et~al.]{van2017neural}
Van Den~Oord, A., Vinyals, O., et~al.
\newblock Neural discrete representation learning.
\newblock \emph{Advances in neural information processing systems}, 30, 2017.

\bibitem[Wang et~al.(2023)Wang, Zhang, HU, Yu, Jin, Gong, and Chen]{wang2023multilevel}
Wang, Z., Zhang, Q., HU, S.-W., Yu, H., Jin, X., Gong, Z., and Chen, H.
\newblock Multi-level protein structure pre-training via prompt learning.
\newblock In \emph{The Eleventh International Conference on Learning Representations}, 2023.
\newblock URL \url{https://openreview.net/forum?id=XGagtiJ8XC}.

\bibitem[Wu et~al.(2024{\natexlab{a}})Wu, Huang, Tan, Gao, Hu, Lin, Liu, and Li]{wu2024psc}
Wu, L., Huang, Y., Tan, C., Gao, Z., Hu, B., Lin, H., Liu, Z., and Li, S.~Z.
\newblock Psc-cpi: Multi-scale protein sequence-structure contrasting for efficient and generalizable compound-protein interaction prediction.
\newblock \emph{arXiv preprint arXiv:2402.08198}, 2024{\natexlab{a}}.

\bibitem[Wu et~al.(2024{\natexlab{b}})Wu, Tian, Huang, Li, Lin, Chawla, and Li]{wu2024mapeppi}
Wu, L., Tian, Y., Huang, Y., Li, S., Lin, H., Chawla, N.~V., and Li, S.
\newblock {MAPE}-{PPI}: Towards effective and efficient protein-protein interaction prediction via microenvironment-aware protein embedding.
\newblock In \emph{The Twelfth International Conference on Learning Representations}, 2024{\natexlab{b}}.
\newblock URL \url{https://openreview.net/forum?id=itGkF993gz}.

\bibitem[Yang et~al.(2020)Yang, Fan, Song, and Lin]{yang2020graph}
Yang, F., Fan, K., Song, D., and Lin, H.
\newblock Graph-based prediction of protein-protein interactions with attributed signed graph embedding.
\newblock \emph{BMC bioinformatics}, 21\penalty0 (1):\penalty0 1--16, 2020.

\bibitem[Yang et~al.(2022)Yang, Zanichelli, and Yeh]{yang2022masked}
Yang, K.~K., Zanichelli, N., and Yeh, H.
\newblock Masked inverse folding with sequence transfer for protein representation learning.
\newblock \emph{bioRxiv}, 2022.

\end{thebibliography}
\bibliographystyle{icml2024}

\clearpage
\renewcommand\thefigure{A\arabic{figure}}
\renewcommand\thetable{A\arabic{table}}
\renewcommand\theequation{A.\arabic{equation}}
\setcounter{table}{0}
\setcounter{figure}{0}
\setcounter{theorem}{0}
\setcounter{equation}{0}
\renewcommand{\dblfloatpagefraction}{.95}

\section*{Appendix}
\subsection*{A. E(3)-equivariant Graph Neural Networks}
Since developing a new E(3)-equivariant architecture is not the focus of this paper, we directly adopt an E(3)-equivariant Graph Neural Network similar to MEAN~\cite{kong2022conditional} for updating node features and coordinates. Suppose the node feature and coordinates of residue $v_i$ in the $l$-th layer are $\mathbf{h}_i^{(l)}$ and $\mathbf{Z}_i^{(l)}$, respectively. We denote the relative coordinates between residue $v_i$ and $v_j$ as $\mathbf{Z}_{i,j}^{(l)}=\mathbf{Z}_{i}^{(l)}\!-\!\mathbf{Z}_{j}^{(l)}$. Then, the message aggregation and updating of the $l$-th layer ($0 \leq l \leq L-1$) for node $v_i$ can be defined as follows
\begin{equation}
\mathbf{m}_{i,j}^{(l)}\!=\!\phi_m\Big(\mathbf{h}_i^{(l)}, \mathbf{h}_j^{(l)}, \frac{(\mathbf{Z}_{i,j}^{(l)})^{\top} \mathbf{Z}_{i,j}^{(l)}}{\big\|(\mathbf{Z}_{i,j}^{(l)})^{\top} \mathbf{Z}_{i,j}^{(l)}\big\|_F}, \mathbf{E}_{i,j}^{(l)}\Big), 
\end{equation}
\begin{equation}
\mathbf{h}_i^{(l+1)}\!=\!\phi_h\Big(\mathbf{h}_i^{(l)}, \sum_{j \in \mathcal{N}(i|\mathcal{E}_{\text{in}})} \mathbf{m}_{i,j}^{(l)}, \sum_{j \in \mathcal{N}(i|\mathcal{E}_{\text{ex}})} \mathbf{m}_{i,j}^{(l)}\Big), 
\end{equation}
\begin{equation}
\mathbf{E}_{i,j}^{(l+1)}\!=\!\phi_e\Big(\mathbf{h}_i^{(l+1)}, \mathbf{E}_{i,j}^{(l)}, \mathbf{h}_j^{(l+1)}\Big),
\end{equation}
\begin{equation}
\hspace{-0.5em}
\begin{small}
\begin{aligned}
\mathbf{Z}_i^{(l+1)}\!=\!\mathbf{Z}_i^{(l)}\!+\!\frac{1}{|\mathcal{N}(i|\mathcal{E}_{\text{in},\text{ex}})|}\!\sum_{j \in \mathcal{N}(i|\mathcal{E}_{\text{in},\text{ex}})}\!\mathbf{Z}_{i,j}^{(l)}\phi_z\big(\mathbf{m}_{i,j}^{(l)}\big).
\end{aligned}
\end{small}
\end{equation}
where $\mathcal{N}(i|\mathcal{E}_{\text{in}})$, $\mathcal{N}(i|\mathcal{E}_{\text{ex}})$, and $\mathcal{N}(i|\mathcal{E}_{\text{in},\text{ex}})$ denote the neighbors of node $v_i$ regarding the internal connections, external connections, and both. Besides, $\phi_m(\cdot)$, $\phi_h(\cdot)$, $\phi_e(\cdot)$, and $\phi_z(\cdot)$ are all implemented as one- or two-layer MLPs with $\operatorname{SiLU}(\cdot)$~\cite{elfwing2018sigmoid} as the activation function. Finally, we output $\widehat{\mathbf{O}}_i\!=\!\mathbf{Z}_i^{(0)}\!-\!\mathbf{Z}_i^{(L)}$ as the prediction on structural noise $\mathbf{O}_i$, i.e., local conformational changes.

\subsection*{B. Huber Loss Function} 
The Huber loss~\cite{huber1992robust} helps to lead to a more stable training procedure, which is defined as follows:
\begin{equation}
l(x, y)=\left\{\begin{array}{l}
0.5(x-y)^2, \text { if }|x-y|<\delta \\
\delta \cdot(|x-y|-0.5 \cdot \delta), \text { else }
\end{array}\right.
\end{equation}
where we set $\delta=1$ in this paper.

\begin{algorithm}[t]
    \caption{Algorithm for the Prompt-DDG}
    \label{algo:1}
    \begin{algorithmic}[1]
  
    \STATE Randomly initializing the parameters of microenvironment encoder, decoder, and prompt codebook $\mathcal{A}$.

    \STATE \# \textit{Prompt Codebook Pre-training}
    \FOR{each iteration}
        \STATE Masking the microenvironment $\mathcal{G}_m$ as $\widetilde{\mathcal{G}}_m$.
        \STATE Encoding microenvironment $\widetilde{\mathcal{G}}_m$ around each mutation $m\in\mathcal{M}$ into representations $\mathbf{h}_m$ by Eq.~(\ref{equ:3});
        \STATE Performing vector quantization on $\mathbf{h}_m$ into prompt codes $z_i$ by the prompt codebook $\mathcal{A}$ by Eq.~(\ref{equ:4});
        \STATE Reconstructing the inputs from prompt embeddings;
        \STATE Optimizing the encoder, decoder, and prompt codebook $\mathcal{A}$ jointly by minimizing the loss of Eq.~(\ref{equ:8}).
	\ENDFOR

    \STATE \# \textit{Prompt-Guided $\Delta\Delta G$ Prediction}
    \STATE Freezing the encoder $f_\theta(\cdot)$ and codebook $\mathcal{A}$, and randomly initializing the parameters of $\Delta\Delta G$ predictor.
    \FOR{each iteration}
        \STATE Combining three prompts of different structural scales by a lightweight prompt adapter by Eq.~(\ref{equ:9}).
        \STATE Add microenvironment-aware prompts to residues in the microenvironment around each mutation.
        \STATE Pooling the structural representations of wild-type and mutant complexes for predicting $\Delta\Delta G$s.
        \STATE Optimizing $\Delta\Delta G$ predictor by minimizing the MSE loss between the predicted and ground-truth $\Delta\Delta G$s.
    \ENDFOR
	\STATE \textbf{return} Trained $\Delta\Delta G$ predictor.
    \end{algorithmic}
\end{algorithm}

\subsection*{C. Comparison with Related Work} 
A recent work, MAPE-PPI~\cite{wu2024mapeppi}, is the first computational approach for microenvironment discovery and encoding. Our Prompt-PDG is partially inspired by it but differs from it in several aspects: (1) MAPE-PPI constructs one \emph{single codebook} characterizing the sequence and structural context of the entire microenvironment, while our Prompt-DDG constructs a \emph{hierarchical codebook} to separately record common microenvironmental patterns at three different structural scales, including 1D residue types, 2D geometric angles, and 3D backbone conformations. (2) MAPE-PPI encodes the \emph{microenvironments around all residues} and utilizes the codebook to generate pre-trained representations for downstream tasks, while our Prompt-DDG encodes only the\emph{ microenvironments around mutations} and generates several prompts for $\Delta\Delta G$ prediction. (3) MAPE-PPI pre-trains the codebook via a \emph{masked codebook} modeling task. However, our Prompt-DDG directly \emph{masks the input microenvironment} (including its residue types, angular statistic, and conformational coordinates), and then trains each sub-codebook with an individual task, aimed at capturing the joint distribution of each residue mutation with three structural scales of the microenvironment.

\subsection*{D. Training Time Complexity Analysis}
The training time complexity of Prompt-DDG comes from four parts: (1) microenvironment encoding $\mathcal{O}(|\mathcal{V}|F^2+|\mathcal{E}|F)$; (2) vector quantization $\mathcal{O}(|\mathcal{M}|\cdot|\mathcal{A}|F)$; (3) prompt combination $\mathcal{O}(|\mathcal{M}|F)$; (4) $\Delta\Delta G$ prediction $\mathcal{O}(|\mathcal{V}|F^2+|\mathcal{E}|F)$, where $|\mathcal{V}|$ and $|\mathcal{E}|$ are the number of nodes and edges, $F$ is the hidden dimension, $|\mathcal{M}|$ is the number of mutations, and $|\mathcal{A}|$ is the size of codebook. The total training time complexity of Prompt-DDG is $\mathcal{O}(|\mathcal{V}|F^2+|\mathcal{E}|F+|\mathcal{M}|\cdot|\mathcal{A}|F)$, which is linear with respect to all $|\mathcal{V}|$, $|\mathcal{E}|$, $|\mathcal{A}|$, and $|\mathcal{M}|$.

\begin{figure*}[!tbp]
	\begin{center}
        \includegraphics[width=0.33\linewidth]{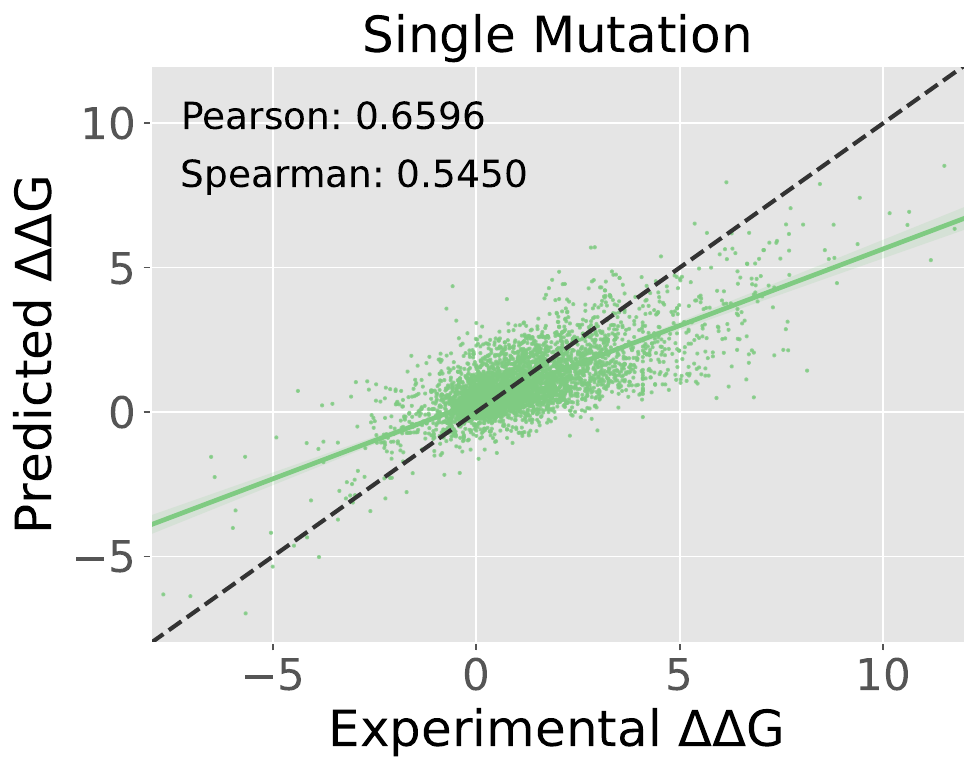}
        \includegraphics[width=0.33\linewidth]{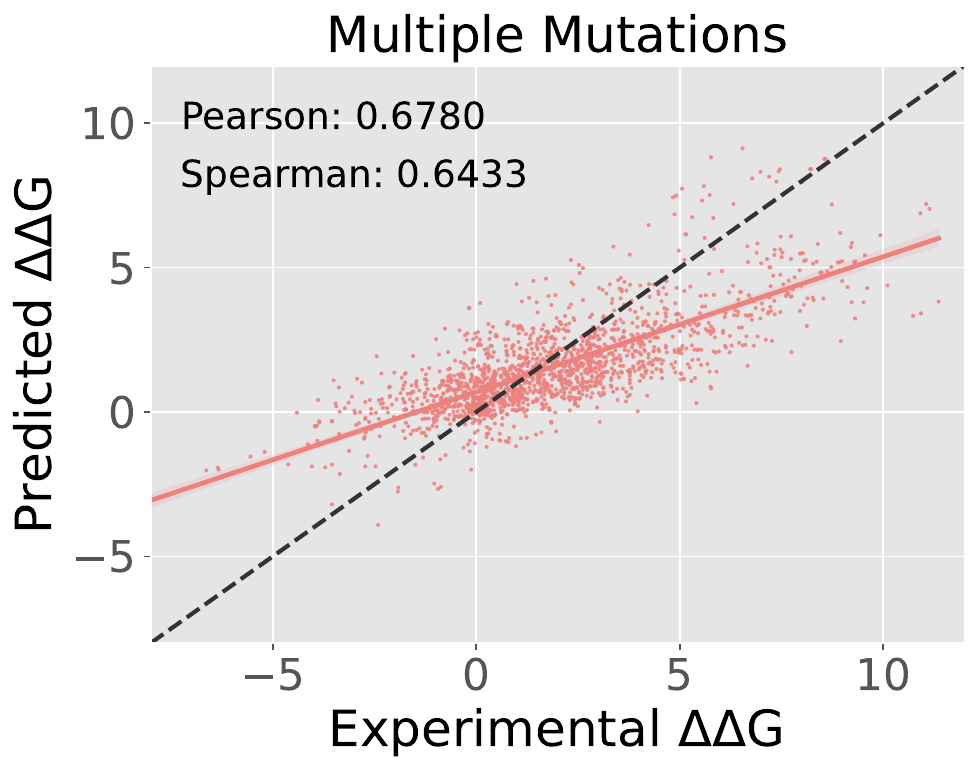}
        \includegraphics[width=0.33\linewidth]{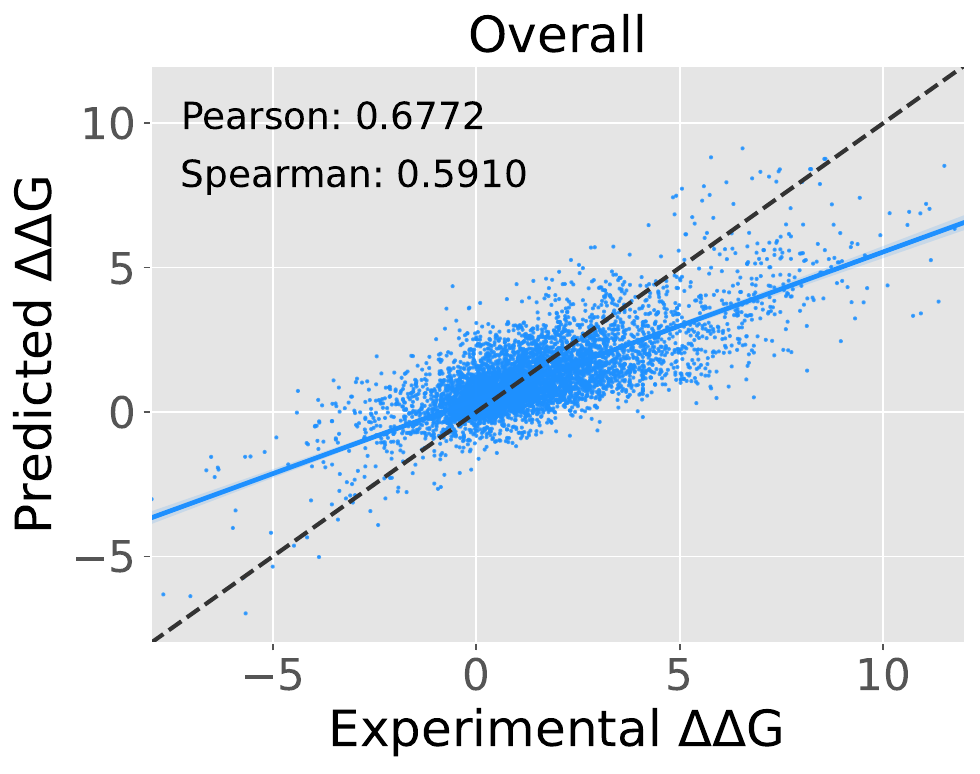}
	\end{center}
	\caption{A visualization comparison of correlations between experimental $\Delta\Delta G$ and $\Delta\Delta G$ predicted by three mutation settings.}
	\label{fig:A1}
\end{figure*}

\subsection*{E. Pseudo-code of Prompt-DDG}
The pseudo-code of the proposed Prompt-DDG framework for $\Delta\Delta G$ prediction is summarized in Algorithm~\ref{algo:1}.

\subsection*{F. Details about Baselines}
The results of all baselines except DiffAffinity in Table.~\ref{tab:1} and Table.~\ref{tab:2} are copied from a previous work~\cite{luo2023rotamer}, which has provided details of various baseline implementations. We refer the interested reader directly to their descriptions in the subsection of \textit{``A.1 Baselines Implementations"}. Besides, the results of DiffAffinity~\cite{liu2023predicting} are taken from their original paper. For a fair comparison, our Prompt-DDG adopts the same self-attention-based network from~\cite{jumper2021highly} as RDE~\cite{luo2023rotamer} does. The only difference is that Prompt-DDG adapts it to graph data by restricting the original global attention computation and message passing to the local microenvironment.

\subsection*{G. Hyperparameters and Implementation Details}
The following hyperparameters are determined by an AutoML toolkit NNI with the hyperparameter search spaces as: Adam optimizer~\cite{kingma2014adam} with $lr=0.0003$, batch size $B=32$, codebook iteration $T_{\text{code}}=2,000$, $\Delta\Delta G$ iteration $T_{\Delta\Delta G}=7,000$, thresholds $d_s=2$, $d_r=15 \AA$, and neighbor number $K=15$ for graph construction, hidden dimension $F=\{128, 256\}$, codebook size $|\mathcal{A}|=\{128, 256, 512\}$, mask ratio $r=\{0.1, 0.2\}$, and loss weights $\eta=0.25$, $\lambda=\{0.01, 0.001\}$. In addition, $\{\phi_{\omega}^{(k)}(\cdot)\}_{k=1}^3$ are implemented as one-layer linear transformation. Besides, We crop structures into patches containing 128 residues by first choosing a seed residue, and then selecting its
127 nearest neighbors based on C-beta distances. 

\subsection*{H. More Correlation Visualizations}
The visualizations of Prompt-DDG for single-point, multi-point, and all-point mutations are provided in Figure.~\ref{fig:A1}.

\end{document}